\documentclass[aps,twocolumn,superscriptaddress]{revtex4-2}
\usepackage{amsfonts}
\usepackage{dcolumn}
\usepackage[left]{lineno}
\usepackage{bm}
\usepackage{tikz}
\usepackage[colorlinks=true,citecolor=blue,urlcolor=blue]{hyperref}
\usepackage{amsmath,amsfonts,amssymb,times,natbib}
\usepackage[standard]{ntheorem}

\begin{document}

\title{Higher-order Topological Parity Anomaly and Half-integer Hall Effect in High-dimensional Synthetic Lattices}

\author{Xian-Hao Wei}

\affiliation{Laboratory of Quantum Information, University of Science and Technology of China, Hefei 230026, China}
\affiliation{Anhui Province Key Laboratory of Quantum Network, University of Science and Technology of China, Hefei 230026, China}
\affiliation{CAS Center For Excellence in Quantum Information and Quantum Physics, University of Science and Technology of China, Hefei 230026, China}

\author{Xi-Wang Luo}\email{luoxw@ustc.edu.cn}
\affiliation{Laboratory of Quantum Information, University of Science and Technology of China, Hefei 230026, China}
\affiliation{Anhui Province Key Laboratory of Quantum Network, University of Science and Technology of China, Hefei 230026, China}
\affiliation{CAS Center For Excellence in Quantum Information and Quantum Physics, University of Science and Technology of China, Hefei 230026, China}
\affiliation{Hefei National Laboratory, University of Science and Technology of China, Hefei 230088, China}

\author{Guang-Can Guo}
\affiliation{Laboratory of Quantum Information, University of Science and Technology of China, Hefei 230026, China}
\affiliation{Anhui Province Key Laboratory of Quantum Network, University of Science and Technology of China, Hefei 230026, China}
\affiliation{CAS Center For Excellence in Quantum Information and Quantum Physics, University of Science and Technology of China, Hefei 230026, China}
\affiliation{Hefei National Laboratory, University of Science and Technology of China, Hefei 230088, China}

\author{Zheng-Wei Zhou}\email{zwzhou@ustc.edu.cn}
\affiliation{Laboratory of Quantum Information, University of Science and Technology of China, Hefei 230026, China}
\affiliation{Anhui Province Key Laboratory of Quantum Network, University of Science and Technology of China, Hefei 230026, China}
\affiliation{CAS Center For Excellence in Quantum Information and Quantum Physics, University of Science and Technology of China, Hefei 230026, China}
\affiliation{Hefei National Laboratory, University of Science and Technology of China, Hefei 230088, China}

\date{\today}

\begin{abstract}
Recent advances in constructing synthetic dimension
provide a powerful tool for exploring exotic topological states of matter in high dimensions.
Here we report that 
the parity anomaly and associated \textit{half-integer} quantized Hall conductance, arising in 2$j$+1 (space-time) dimensions
with a single or odd number of Dirac cones, can be realized by the boundary states of $n$-th order topological insulators in (2$j$+$n$)-dimensional synthetic lattices. 
We establish a general bulk-boundary correspondence by integrating the
``nested" Wilson loop theory with the time-reversal polarization
at highly-symmetric momenta, a set of $Z_2$ topological invariants are extracted which determines the number of higher-order-boundary Dirac cones and their locations. 
We develop a general construction procedure for Hamiltonians supporting such higher-order topological parity anomaly.
Moreover, we propose an experimental implementation scheme based on photonic synthetic dimensions and provide a method for probing the associated half-integer Hall conductance by the transmission spectra.
Our work offers the realization and characterization of parity anomaly in general high-dimensional higher-order topological insulators and open an avenue for exploring fundamental physics and possible device applications enabled by manipulating Dirac cones.

\end{abstract}

\maketitle
\textbf{Introduction.} Dimension is a fundamental concept in physics and 
critically determines the topological characteristics of quantum systems~\cite{kanereview,shouchengreview}, and higher-dimensional systems may host more exotic topological phases beyond lower-dimensional analogs. This profound influence becomes particularly evident when one contrasts
the 2-dimensional (2D) topological insulator (TI)~\cite{kanez2,haldanemodel} exhibiting integer Hall effect through 1D boundary modes, with the 3D TI~\cite{threedimensionliangfu} possessing single Dirac cone on each 2D boundary that could cause
\textit{half-integer} quantized Hall conductance under time-reversal symmetry-breaking perturbations---a manifestation of the parity anomaly in quantum field theory~\cite{axionexp2022,parityanomalyorigin1,parityanomalyorigin2,parityanomalyorigin3,parityanomalyorigin4,parityanomalyshen,parityanomalyvalley1}. Because of which,
the 3D TI also serves as axion insulator~\cite{photonicaxion,axionexp2018} and has garnered 
significant attention. Generally, parity anomaly and associated half-integer Hall response
may emerge in any 2$j$-dimensional systems with single or odd number of Dirac cones~\cite{parityanomalyorigin1,parityanomalyorigin2}. 
Recent advances in constructing synthetic lattices utilizing atomic or photonic internal degrees of freedom~\cite{celisyntheticdimension,coldatomexp1,coldatomexp2,yanboexp,gadwayexp,dwwangscience,luo2015,suwang,photoexp20,photosynreview1,photosynreview2,photosynreview3,yuantwosynthetic,photoexp1,photoexp4,photoexp15,photoexp19} provide the possibility of studying the general parity anomaly and other more exotic topological phases (such as 4D quantum Hall effect~\cite{4dhallhujiangping,qixiaoliang1,4datomhalltheo,4datomnew,4dfloquet,4dphotonichalltheo,4dpumpatom,4dpumpphoton}, 5D Weyl surfaces~\cite{5dweyltopophasetransition,meta5dsurface}, tensor and Yang monopoles~\cite{tensormonopole1,tensormonopole2,5dyangmonopole}, among others~\cite{4dtopoanderson,zhuyanqing1}) in dimensions beyond three.

Meanwhile, the exploration of topological states of matter has extended to higher-order systems supporting novel boundary states on corners or hinges with unconventional bulk-boundary correspondence~\cite{higherorderscience,higherorderprb,higherorderxureview,higherordernrp}. In general, a $d$-dimensional $n$-th order TI can host ($d-n$)-dimensional  boundary modes. 
To date, despite extensive theoretical and experimental investigations~\cite{higherorder2013,higherorderscienceadvances,higherorderxiwang,higherorderfangchen,higherordererjiefenlei,higherorderexact,higherorder170page,higherorderexpextra4,higherorderhybrid,higherorderdirac1,higherorderhybridexp,higherorderanderson,higherordermeifeng,higherordersyntheticdimension1,higherordersyntheticdimension2}, most research on higher-order TIs has revolved around the boundary modes on 1D hinge and 0D corner in 3D and 2D systems, leaving higher-dimensional generalizations largely unexplored due to the reliance on additional dimensions exceeding the capabilities of solid-state materials.
This raises the following fundamental questions:
Whether the general parity anomaly---manifested as a 2$j$-D Dirac cone and half-integer Hall response---occurs in general higher-order TIs? If so, what is the bulk-boundary correspondence and how to realize it experimentally?

In this article, we address these important questions by 
utilizing synthetic lattices.
We first take a 4D lattice model as an example and demonstrate that the 2D parity anomaly can be realized by the boundary states of the system in the second-order topological phase. Each corner can host a single Dirac cone that contributes half-integer Hall conductance under time-reversal breaking corner perturbations, as confirmed by layer-resolved Chern numbers. Then, we establish a general bulk-boundary correspondence by integrating the
``nested" Wilson loop theory with the time-reversal polarization
at highly-symmetric momenta and develop a construction procedure for general $d$-dimensional $n$-th order topological Hamiltonians supporting general parity anomaly. A set of $Z_2$ topological invariants, that determine the number of Dirac cones on the corresponding corner and their positions in momentum space, are innovatively introduced, which is applicable to any systems with time-reversal and reflection symmetric bulk.
Finally, we propose experimental realization schemes based on coupled cavity arrays with photonic internal degrees of freedom (e.g., orbital angular momentum, frequency, etc.) as additional synthetic dimensions, we suggest that the half-integer Hall conductance can be measured by the transmission spectra 
which capture the {mode density} and allow the extraction of layer-resolved Chern numbers via the Streda formula.\\

\begin{figure}[t]
\includegraphics[width=1.0\linewidth]{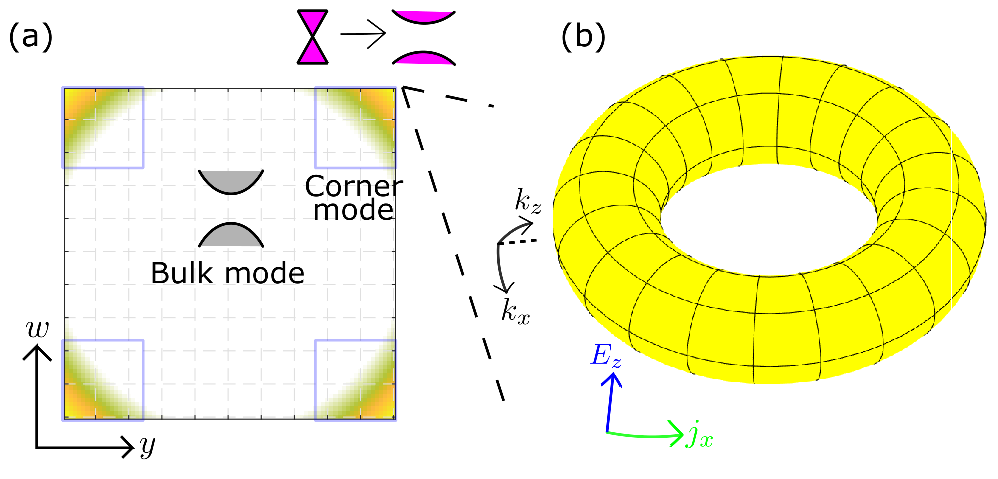}
\caption{Schematic of 4D second-order topological insulator for the model Eq.~(\ref{Eqhamiltonian1}), with open (periodic) boundary condition along $y,w$ ($x,z$). The topology of gapped bulk is responsible to the gapless Dirac cone at each corner, time-reversal breaking perturbation can lead to a corner gap and half-integer quantized Hall conductance.}
\label{fighamitonian}
\end{figure}

\textbf{Model.} We first consider a 4D eight-band lattice model with Bloch Hamiltonian:
\begin{eqnarray}
    H(\mathbf{k})
    &=&\lambda_\text{SO} (\sin k_x\Gamma_1+\sin k_y\Gamma_2+\sin k_z\Gamma_3+\sin k_w\Gamma_4) \nonumber\\
    &&+(m_1+2t_x\cos k_x+2t_y\cos k_y)\Gamma_5\nonumber\\
    &&+(m_2+2t_z\cos k_z+2t_w\cos k_w)\Gamma_6, \label{Eqhamiltonian1} 
\end{eqnarray}
where $\Gamma_j$ are the Dirac matrices satisfying Clifford algebra $\{\Gamma_j,\Gamma_{j'}\}=2\delta_{jj'}$. 
For our eight-band model with 3 Paulis, the Dirac matrices read (different choices are equivalent): 
$\{\Gamma_{(1)-(7)}=(\sigma_z\tau_zs_x,\sigma_y,\sigma_z\tau_zs_y,\sigma_z\tau_y,\sigma_x,\sigma_z\tau_x,\sigma_z\tau_zs_z)\}$
with $\sigma_j$, $\tau_j$, $s_j$ ($j=x,y,z$) the Pauli matrices representing different (pseudo-)spins or sub-lattice sites. 
The first line on the right-hand side of Eq.~(\ref{Eqhamiltonian1}) can be viewed as the spin-orbit coupling, while the last two lines as kinetic energy. Here we consider isotropic tunneling $t_x=t_y=t_z=t_w\equiv t$ for simplicity and set $2t=1$ as the energy unit.
The eigenenergies are: $E(\mathbf{k})=\pm\sqrt{\sum_jh_j(\mathbf{k})^2}$ with $h_j(\mathbf{k})$ the coefficient of $\Gamma_j$ in the Hamiltonian, each of the two bands has four-fold degeneracy, the band gap closes only at the 9 points on $m_1$-$m_2$ plane with $m_1=0,\pm2$ and $m_2= 0,\pm2$. Note that these gap closing points are independent from $\lambda_\text{SO}$, without loss of generality, we will set $\lambda_\text{SO}=1$ in the following.
Our model preserves time-reversal and reflection symmetries $TH(\mathbf{k})T^{\dagger}=H(-\mathbf{k})$ with $T=s_yK$ ($K$ is the complex conjugation), and $M_iH(\mathbf{k})M_i=H(-k_i,k_{j\neq i})$ with $M_x=s_y$, $M_y=\sigma_x\tau_zs_z$, $M_z=s_x$, $M_w=\tau_xs_z$. 
For isotropic tunneling, the model also has $C_4$ rotational symmetries in the $x$-$y$ and $z$-$w$ planes with $C_4^{x,y}=\frac{I-\Gamma_1\Gamma_2}{\sqrt{2}}$ and  $C_4^{z,w}=\frac{I-\Gamma_3\Gamma_4}{\sqrt{2}}$, as well as a combined rotational symmetry $C_4^{xy,zw}H(k_x,k_y,k_z,k_w)(C_4^{xy,zw})^{-1}=H(-k_z,-k_w,k_x,k_y)$
when $m_1=m_2$, with $C_4^{xy,zw}=\frac{I-\Gamma_1\Gamma_3}{\sqrt{2}}\frac{I-\Gamma_2\Gamma_4}{\sqrt{2}}\frac{\Gamma_5+\Gamma_6}{\sqrt{2}}$ (similar for $m_1=-m_2$).

\begin{figure}[t]
\includegraphics[width=1.0\linewidth]{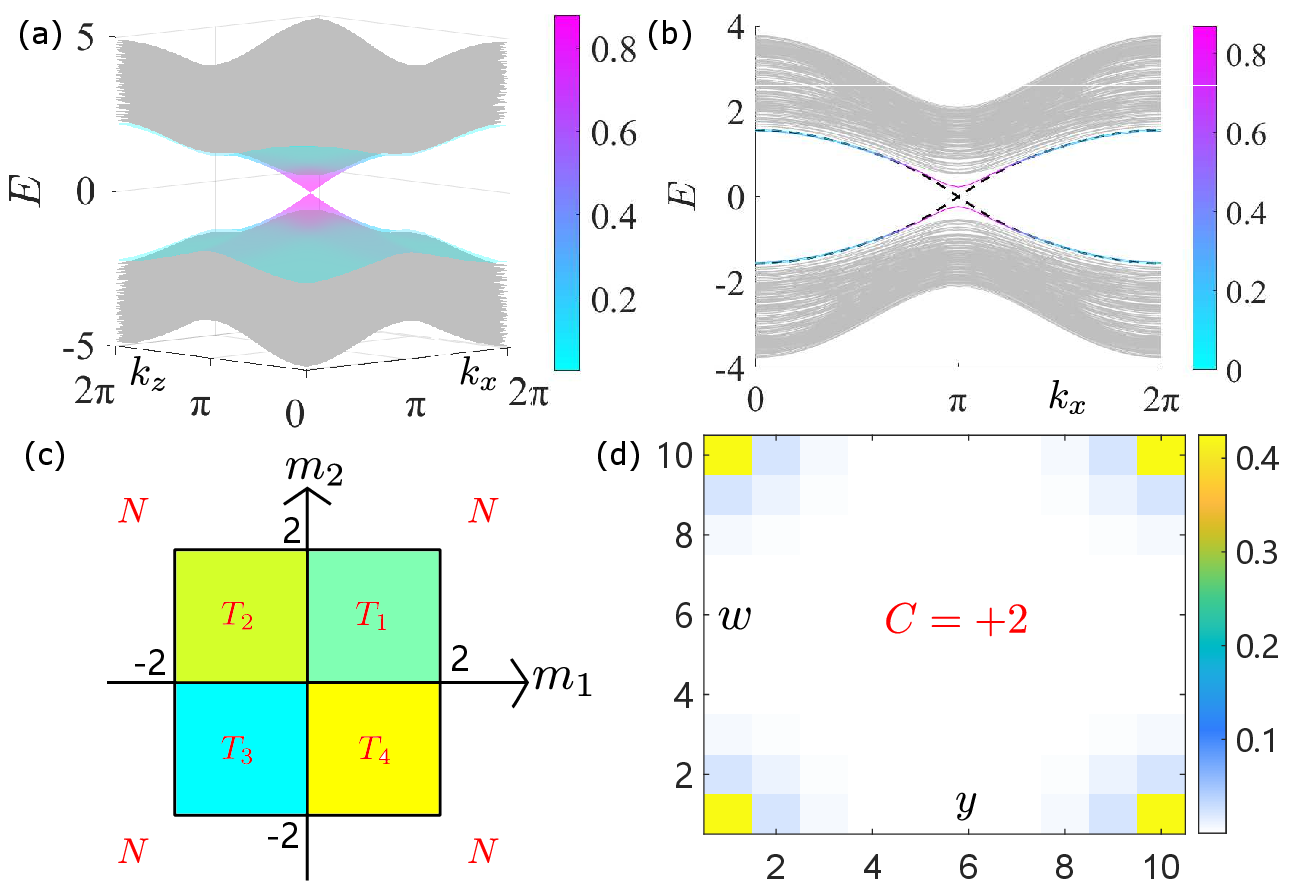}
\caption{(a) Bulk (gray) and corner (colored) bands of the model Eq.~(\ref{Eqhamiltonian1}) with boundary condition shown in Fig.~\ref{fighamitonian}.
Color bar shows the populations on the four corner sites.
(b) Energy bands at $k_z=\pi$ with additional mass term $H_\text{M}=-Ms_z$ on the four corner sites. Dashed lines represent the gappless corner modes in (a).
(c) Phase diagram in the $m_1$-$m_2$ plane, with $N$ the normal phase and $T_i$ the topological phases.
(d) Layer-resolved Chern number with half-quantized corner contribution. We set $m_1=m_2=1.5$ with system size $N_y=N_w=10,N_x=N_z=36$ in (a,b,d) and $M=0.4$ in (b,d).}
\label{figphasetransition}
\end{figure}

We find the model defines a 4D second-order topological insulator with massless Dirac cones formed by topological corner modes, one for each corner, leading to higher-order topological parity anomaly. To show this, we consider
open-boundary (period-boundary) conditions along $y$ and $w$ ($x$ and $z$) directions, as shown in Fig.~\ref{fighamitonian}. The corner modes can be obtained by assuming the ansartz~\cite{higherorderexact} $|\Psi_\text{corner}\rangle\propto\sum_{y,w\geq0}(\kappa_1)^y(\kappa_2)^w|y,w\rangle\otimes|\xi\rangle$ and solving 
the eigen equation $H|\Psi_\text{corner}\rangle=E_\text{corner}|\Psi_\text{corner}\rangle$, with $|\xi\rangle$ the spin state, see SI.~I.~A (i.e., section I.~A in Supplementary Information) for more details. For the corner around $y,w=0$, the two solutions read $|\xi\rangle=\{\left|\uparrow\uparrow\uparrow\right\rangle_{\sigma\tau s},\left|\uparrow\uparrow\downarrow\right\rangle_{\sigma\tau s}\}$, with $\kappa_1=-m_1-\cos k_x,\kappa_2=-m_2-\cos k_z$. Within the subspace spanned by the two corner modes, the effective Hamiltonian can be written as
\begin{eqnarray}
    H_\text{corner}(k_x,k_z)=s_x\sin k_x+s_y\sin k_z\label{Eqeffectivecorner}    
\end{eqnarray}
which forms a Dirac cone in $x$-$z$ space as illustrated in Fig.~\ref{fighamitonian}. 
Similar results can be obtained for other three corners, the numerical band structure are shown in
Fig.~\ref{figphasetransition}(a), where the gapped bulk corresponds to the gray bands, while the corner Dirac cones are given by the colored bands with color bar showing the populations on the four corner sites.

The above corner modes (around $y,w=0$) require that $|\kappa_1|,|\kappa_2|<1$, so the existence and momentum position of the Dirac cone depend on $m_1,m_2$. In particular, the corner Dirac cone exists and the system is topological in the region
$|m_1|,|m_2|<2$, and is trivial otherwise. The Dirac cone is located at $k_x=\pi,0$ ($k_z=\pi,0$) for positive and negative $m_1$ ($m_2$), respectively, and thus the system has 4 topological phases as shown in Fig.~\ref{figphasetransition}(b). Recall that the bulk gap closes only at certain points in the $m_1$-$m_2$ plane. Thus, instead of the bulk bands, the edge spectra (states located on $x=0$ and $z=0$ edges) close the gap across the phase boundary and lead to the phase transition of the corner modes, a typical phenomenon for higher-order TIs~\cite{higherorderscience,higherorderprb,higherorderxureview,higherordernrp}.

By adding
$T$-breaking perturbation mass term  (analogous to gauge-invariant regularization process in quantum field theory), the parity symmetry (i.e., reflection with respect to the Dirac point) is also broken, leading to half-integer Hall conductivity~\cite{parityanomalyorigin1,parityanomalyorigin2,parityanomalyorigin3,parityanomalyorigin4,parityanomalyshen,axionexp2022}. We introduce the mass term by a Zeeman field $H_\text{M}=-Ms_z$ applied on the four corner sites in $y$-$w$ space (all sites in $x$-$z$ space), it opens a gap in the corner Dirac cone as shown in Fig.~\ref{figphasetransition}(c). Then, each corner contributes $1/2$-quantized Hall current $j_x=\frac{1}{2}\frac{e^2}{h}E_z$ along $x$ in the presence of the gradient field $E_z$ along $z$, a typical consequence of parity anomaly, as illustrated in Fig.~\ref{fighamitonian}. This is verified by calculating the layer-resolved Chern number~\cite{localchern1,localchern2} $C(y,w)=\frac{1}{2\pi}\int dk_xdk_z \text{Tr}[\Omega_{k_x,k_z}\rho_{k_x,k_z}(y,w)]
$, with $\Omega_{k_x,k_z}$ and $\rho_{k_x,k_z}(y,w)$ the non-Abelian Berry connection and local density-of-state matrix of the occupied bands (as elaborated in SI.~I.~B and C).
We can see that the Hall conductance of each corner state is half-quantized, as shown in Fig.~\ref{figphasetransition}(d). The total Chern number is always quantied to integers. We have $C=2$ and $-2$ in phases $T_{2,4}$ and $T_{1,3}$, respectively (since their Dirac-cone chiralities are different). It is worthy noticing that the sign of the mass and the direction of the Hall current can be controlled independently for each corner. For general open boundaries along 
directions $q_1,q_2$, the above results of the corner Dirac cones also apply if
$q_1\in (x,y)$ and $q_2\in (z,w)$, as elaborated in SI.~I.~D.

\begin{figure}[t]
\includegraphics[width=1.0\linewidth]{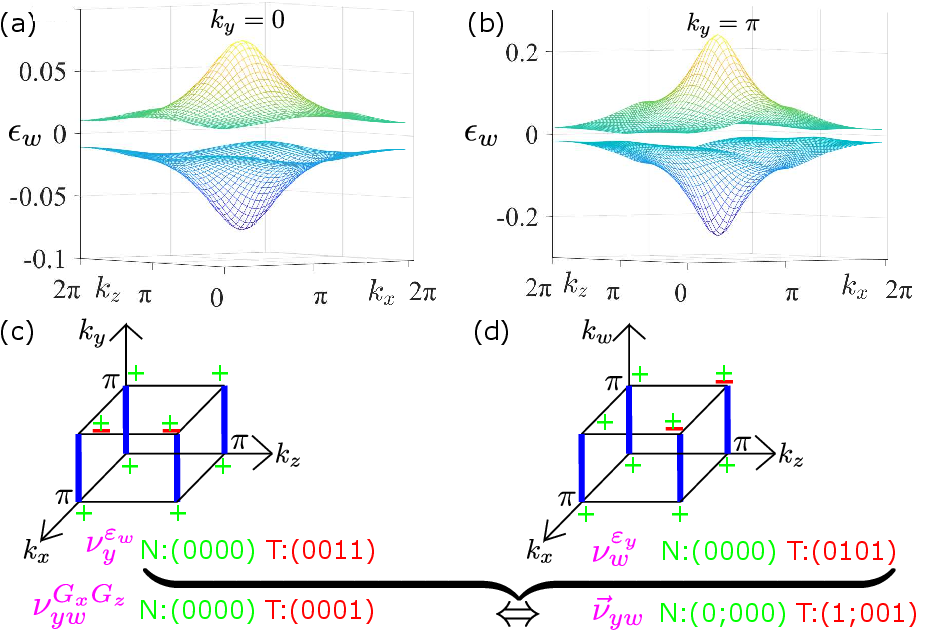}
\caption{[(a) and (b)] The Wannier band structures $\varepsilon_{w,\mathbf{k}}$ with $k_y=0$ and $k_y=\pi$, respectively. $m_1=m_2=1.5$. [(c) and (d)] The parity of the Wannier-band basis (i.e., eigenvalue of the reflection operator) $\eta^{\varepsilon_{w}}(\textbf{G})$ and $\eta^{\varepsilon_{y}}(\textbf{G})$, respectively, for Wannier sector $\varepsilon\in(0,1/2)$. The time-reversal polarizations $\nu_y^{\varepsilon_w}(G_x,G_z)$ and $\nu_w^{\varepsilon_y}(G_x,G_z)$, as well as topological invariants 
$\nu_{yw}^{G_x,G_z}$ and $\vec{\nu}_{yw}$ are also shown for both the normal and topological phases.}
\label{mainfigcharacterize}
\end{figure}

\textbf{{Bulk-boundary correspondence.}} To show the topological originating of the corner parity anomaly, we combine the nested Wilson loop method~\cite{higherorderscience,higherorderprb} with the time-reversal polarization (a $Z_2$ invariant characterizing topological insulator)~\cite{fuliangz2,fuliangz2withinversion}, 
which give rise to a series of $Z_2$ topological invariants determining the number of Dirac cones and their locations. For the 4D lattice model of Eq.~(\ref{Eqhamiltonian1}), we introduce the 4D Wilson loop operator
\begin{equation}
    W_{w,\mathbf{k}}=\mathcal{P}\exp[-i\int_{0}^{2\pi}A_w(\mathbf{k})dk_w]
\end{equation}
where $A_w(\mathbf{k})=i\langle u_{m',\mathbf{k}}|\partial_{k_w}|u_{m,\mathbf{k}}\rangle$ is the non-Abelian Berry connection, $|u_{m,\mathbf{k}}\rangle$ is the $m$-th occupied Bloch band satisfying $H(\mathbf{k})|u_{m,\mathbf{k}}\rangle=E_m(\mathbf{k})|u_{m,\mathbf{k}}\rangle$ with $\langle u_{n,\mathbf{k}}|u_{m,\mathbf{k}}\rangle=\delta_{n,m}$. $\mathcal{P}$ is the path-ordering operator.
The Wilson loop operator leads to the Wannier Hamiltonian $H_{W_w}(\mathbf{k})=-i/2\pi \log  W_{w,\mathbf{k}}$ with $H_{W_w}(\mathbf{k})|\varepsilon_{w,j,\mathbf{k}}\rangle=\varepsilon_{w,j,\mathbf{k}}|\varepsilon_{w,j,\mathbf{k}}\rangle$ and $j$ the Wannier band index. $|\varepsilon_{w,j,\mathbf{k}}\rangle$ satisfies $\langle \varepsilon_{w,j',\mathbf{k}}|\varepsilon_{w,j,\mathbf{k}}\rangle=\delta_{j'j}$ and $\varepsilon_{w,j,\mathbf{k}}$ obeys the identification $\varepsilon_{w,j,\mathbf{k}}=\varepsilon_{w,j,\mathbf{k}} \mod 1$. Due to the reflection symmetry, the Wannier bands appear in $\pm \varepsilon$ pairs, or locked at the symmetric values $0,1/2$ (as demonstrated in SI.~II.~A), and gapped Wannier bands can carry topological invariants. 

The typical Wannier bands in the topological phase are shown in Figs.~\ref{mainfigcharacterize} [(a) and (b)], with a gap at $\varepsilon=0$. We will focus on the Wannier sector with $\varepsilon_j\in (0,1/2)$ with two degenerate Wanner bands, and define the Wannier-band basis $|\phi_{j,w}(\mathbf{k})\rangle=\sum_{m=1}^{N_\text{occ}}[|\varepsilon_{w,j,\mathbf{k}}\rangle]_m|u_{m,\mathbf{k}}\rangle$ with $N_\text{occ}$ the number of occupied energy bands and $\langle \phi_{j,w}(\mathbf{k})|\phi_{j',w} (\mathbf{k})\rangle=\delta_{j,j'}$~\cite{higherorderscience}. 
In the presence of reflection symmetry, $|\phi_{j,w}(\mathbf{G})\rangle$ satisfies $-iM_xM_yM_z|\phi_{j,w}(\mathbf{G})\rangle={\eta}^{\varepsilon_w}(\mathbf{G})|\phi_{j,w}(\mathbf{G})\rangle$, where $\mathbf{G}$ refers to the time-reversal invariant momenta (i.e., $G_{x,y,z}=0,\pi$).
Within each Wannier sector, the Wannier-band basis form time-reversal pairs and share the same eigenvalue $\eta^{\varepsilon_w}=\pm1$ (as proved in SI.~II.~A). There are eight $\eta^{\varepsilon_w}$'s corresponding to the eight $\mathbf{G}$ momenta (since $k_w$ is integrated out), as shown in Figs.~\ref{mainfigcharacterize} [(c) and (d)]. Now we define the nested time-reversal polarization $\nu_y^{\varepsilon_w}(G_x,G_z)$ along $y$ direction  through
\begin{equation}
(-1)^{\nu_y^{\varepsilon_w}(G_x,G_z)} =\prod_{G_y=0,\pi}\eta^{\varepsilon_w}(\mathbf{G}),
\end{equation}
here ``nested" means it is based on the Wannier-band basis.
Notice that $\eta^{\varepsilon_w}(\mathbf{G})$ is associated to the 
Pfaffian $\eta^{\varepsilon_w}(\mathbf{G}) ={\text{Pf}[S^{\varepsilon_w}(\mathbf{G})]}/{\sqrt{\text{Det}[S^{\varepsilon_w}(\mathbf{G})]}}$ with skew matrix $S^{\varepsilon_w}_{j'j}(\mathbf{G})\equiv\langle\phi_{j',w}(\mathbf{G})|T|\phi_{j,w}(\mathbf{G})\rangle$ under transverse gauge choice~\cite{fuliangz2}.
Similarly, we can calculate the Wannier bands along $y$ and obtain the nested time-reversal polarization ${\nu_w^{\varepsilon_y}(G_x,G_z)}$ along $w$. Finally, we arrive at the $Z_2$ topological invariant 
\begin{equation}
{\nu_{yw}^{G_xG_z}} = \nu_y^{\varepsilon_w}(G_x,G_z)\nu_w^{\varepsilon_y}(G_x,G_z).
\end{equation}
$\nu_{yw}^{G_xG_z}=1$ or 0 indicates the existence or absence of a Dirac cone at $k_x=G_x,k_z=G_z$, for each corner of the $y$-$w$ space with open boundaries (see more details in SI.~II.~B). 

There are four independent $Z_2$ topological invariants $\nu_{yw}^{00}$, $\nu_{yw}^{0\pi}$, $\nu_{yw}^{\pi\pi}$, $\nu_{yw}^{\pi0}$, and the even-odd property of the total Dirac cones at each corner is characterized by $\nu_{yw}^\text{total}=\sum_{G_x,G_z} \nu_{yw}^{G_x,G_z} \mod 2$. 
Similar to 3D TIs~\cite{threedimensionliangfu}, here we use  $\vec{\nu}_{yw}= (\nu_{yw}^\text{total}; \nu_{yw}^{00}, \nu_{yw}^{0\pi},\nu_{yw}^{\pi\pi})$.
$\nu_{yw}^\text{total}=1$ implies the realization of higher-order topological parity anomaly with odd number of Dirac cones at each corner, while $\nu_{yw}^{00}, \nu_{yw}^{0\pi},\nu_{yw}^{\pi\pi}$ encode the location information of the Dirac cones. 
For example, $\vec{\nu}_{yw}={(0;0,0,0)}$ in the trivial phase and $\vec{\nu}_{yw}={(1;0,0,1)}$ in the topological $T_1$ phase, as shown in Fig.~\ref{mainfigcharacterize}.
Across the phase boundary, the Wannier band gap closes and reopens, leading to the change of topological invariants. 
It is straightforward to 
obtain the topological invariants $\vec{\nu}_{q_1q_2}$ for open boundaries along different directions.

\begin{figure}[t]
\includegraphics[width=1.0\linewidth]{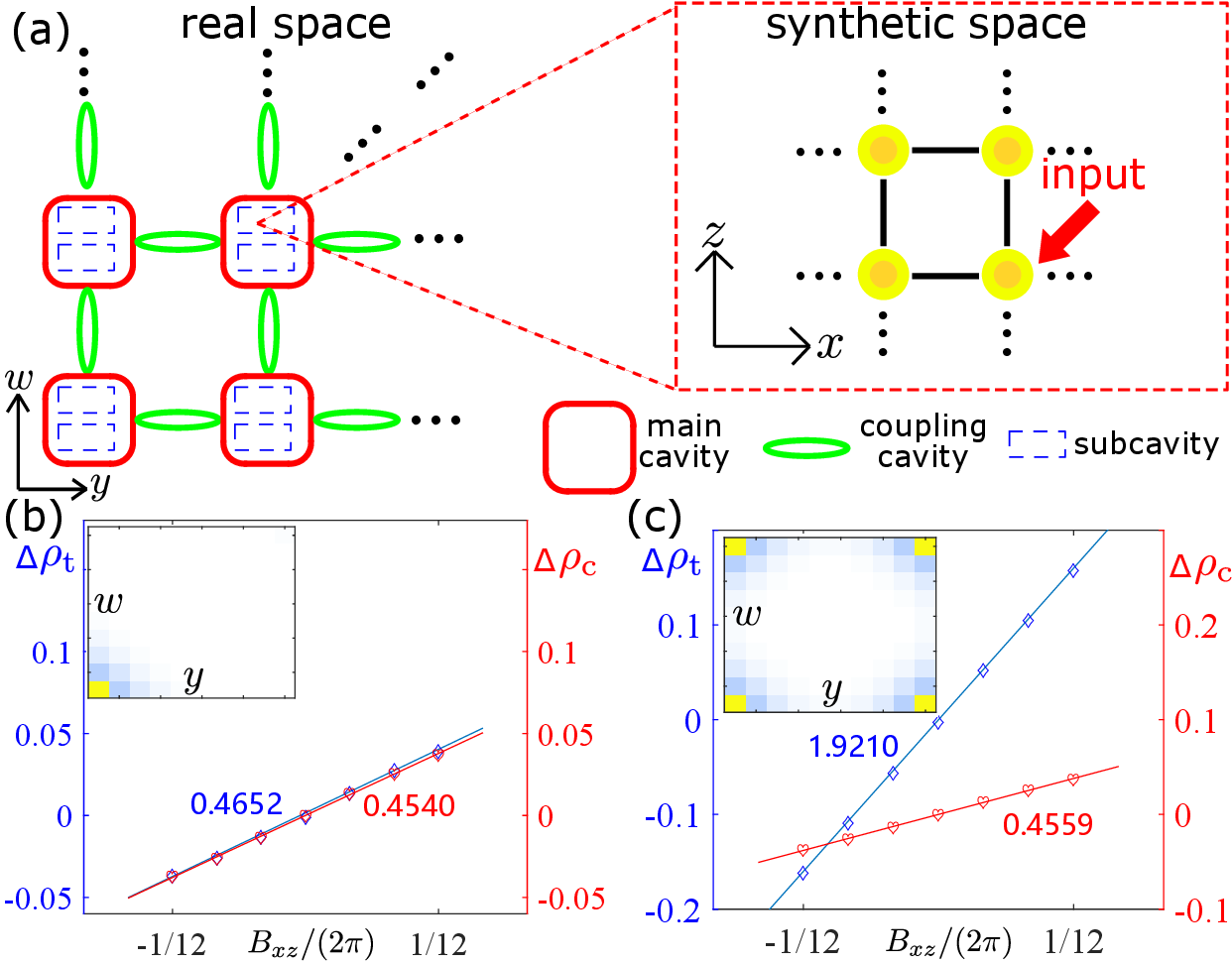}
\caption{(a) Experimental implementation of our Model Eq.~(\ref{Eqhamiltonian1}) based on 2D coupled cavity array with two
additional synthetic dimensions. [(b) and (c)] Numerical simulation of the mode density through photon transmission for different $B_{xz}$, the slopes (marked by the digits) represent the corresponding Chern number.  
$\Delta\rho_\text{t}=\sum_{y_o,w_o} [\rho(y_0,w_0,B_{xz})-\rho(y_0,w_0,0)]$ is the total mode-density derivation (diamonds), similarly for the corner mode-density derivation (hearts) $\Delta\rho_\text{c}$, except the summation is taken over a single corner region. The insets show the extracted layer resolved Chern number. The mass term is applied on one corner in (b) and four corners in (c). Other parameters are: $m_1=m_2=1.5$, $M=0.4$ and $\gamma=0.01$.}
\label{mainsyn}
\end{figure}

\textbf{{Generalization.}} We now explore the possibility of generalizing our results and provide a general procedure for constructing $d$-dimensional $n$-th order topological insulators, which host ($d-n$)-dimensional Dirac cone as topological boundary modes. 
First, we need to generate $d+n$ Dirac matrices.
Using 
$p=\lceil\frac{d+n-1}{2}\rceil$ Paulis, we
can construct the Dirac-matrix set
$\Gamma^{(p)}=\{\sigma_x^{(p)},\sigma_y^{(p)},\sigma_z^{(p)}\otimes\Gamma^{(p-1)}\}$ with
$\Gamma^{(1)}=\{\sigma_x^{(1)},\sigma_y^{(1)},\sigma_z^{(1)}\}$,
so $\Gamma^{(p)}=\{\Gamma_1,\cdots,\Gamma_{d+n},\cdots\}$ contains at least $d+n$ elements. Then, we divide the $d$ dimensions $X=\{x_1,\cdots,x_d\}$ into $n$ set $X=\{X_1,\cdots,X_n\}$, each set $X_j$ contains $d_j>0$ dimensions with $\sum_j d_j=d$.
The general
Hamiltonian reads:
\begin{equation}
    H(\mathbf{k})=\sum_{j=1}^{d}\sin k_j\Gamma_j+\sum_{j=1}^{n}\left(m_j+\sum_{l\in X_j}\cos k_l\right)\Gamma_{j+d}
    \label{generalhk}
\end{equation}
The system is topological in the regime $|m_j|<d_j$, supporting ($d-n$)-dimensional Dirac cone at the intersection corner under open boundaries of $q_1,\cdots,q_n$ with $q_j\in X_j$ (as demonstrated in SI.~III.~A). When $(d-n)$ is even (odd), the system realizes a general higher-order topological parity (chiral) anomaly. With proper mass term to open corner gaps, anomalous Hall conductivity---characterized by $(d-n)/2$-th Chern number---arises (see SI.~III.~B for more details).

It is straightforward to extend our bulk-boundary correspondence to the general case of higher-order topological parity anomaly. The nested time-reversal polarization $\nu_{q_j}^{\varepsilon_{\bar{q}_j}}$ along $q_j$ can be calculated using the Wannier-band basis after evaluating the nested Wilson loops along $\bar{q}_j=(q_1,\cdots,q_{j-1},q_{j+1},q_n)$, the $Z_2$ topological invariants then read $\nu_{q_1\cdots q_n}=\prod_j \nu_{q_j}^{\varepsilon_{\bar{q}_j}}$. We have $2^{d-n}$ 
$Z_2$ invariants associated with the symmetric momenta in the $d-n$ dimensions. 
In fact, our topological characterization relies solely on time-reversal and reflection symmetries (with $C_4$ symmetry non-essential), making it applicable to any models preserving these symmetries, including 3D second-order hinge states~\cite{higherorderscienceadvances} and higher-order chiral anomaly with odd $(d-n)$. 
In SI.~III.~C and D, we have considered another 4D model and a 6D model, respectively, and demonstrated the existence of second-order parity anomaly and bulk-boundary correspondence.

\textbf{{Experimental consideration.}} It is possible to implement the lattice model Eq.~\ref{Eqhamiltonian1} using synthetic lattices in atomic, electric-circuit, or photonic systems~\cite{photosynreview1,higherorderanderson}.
Here we focus on the realization with photonic systems due to their unlimited internal degrees of freedom~\cite{photosynreview1,photosynreview2,photosynreview3} [e.g., orbital angular momentum (OAM) and frequency]. 
We employ a 2D coupled cavity array in real space, together with two synthetic dimensions based on frequency and/or OAM to simulate the 4D lattice (photons with different OAM and frequencies can resonate simultaneously inside the main cavity~\cite{yuantwosynthetic}), as shown in Fig.~\ref{mainsyn} (a). 
The (pseudo-)spin states can be realized by  
two polarization modes ($\sigma$), clockwise and counterclockwise modes ($\tau$), and two sub-cavities ($s$).
The spin-orbit coupling and kinetic tunneling can be realized by properly designed 
auxiliary coupling cavities. For example, the tunneling phase 
can be controlled by the length-imbalance of the two arms in the coupling cavity~\cite{luo2015}, so polarization-dependent phase leads to the $\sigma_z$-type coupling. Polarization-flip ($\sigma_x$-type) coupling can be realized by inserting a $\lambda/4$-wave-plate into the coupling cavity. For the clockwise-counterclockwise modes and sub-cavities, the coupling can be realized similarly~\cite{photoexp1}, combining together one can obtain couplings for all $\Gamma_j$ matrices (note that $\sigma_y=i\sigma_x\sigma_z$). 
Furthermore, multi-synthetic dimensions can be realized using solely OAM or frequency, enabled by long-range tunnelings~\cite{suwang,photoexp20}. Therefore, photonic synthetic lattices not only enable the realization of our 4D model with high feasibility but also provide a suitable platform for exploring physics in even higher dimensions.

To probe the parity anomaly, we propose to measure 
the anomalous half-integer Hall conductivity in the presence of mass term $H_\text{M}$. The properties of the total Hamiltonian are fully captured by the photonic transmission operator~\cite{luo2015,yuantwosynthetic} ${\mathbb{T}}_{\mathbf{r}_\text{o},\mathbf{r}_\text{i}}(\omega)=\langle \mathbf{r}_\text{o}|\frac{-i\gamma}{\omega-{H}+i\gamma}|\mathbf{r}_\text{i}\rangle$, $\mathbf{r}=(x,y,z,w,\alpha)$ is the site-spin index, $\gamma$ is the photonic decay rate. We fix the input index $(x_\text{i},z_\text{i})$ and measure the total intensity transmission to sites $(y_\text{o},w_\text{o})$, that is $\mathcal{T}_{y_\text{o}w_\text{o}}(\omega)=\sum_{\mathbf{r}'_\text{o},\mathbf{r}''_\text{i}}|{\mathbb{T}}_{\mathbf{r}_\text{o},\mathbf{r}_\text{i}}(\omega)|^2$
with $\mathbf{r}'_\text{o}=(x_\text{o},z_\text{o},\alpha_\text{o})$ and $\mathbf{r}''_\text{i}=(y_\text{i},w_\text{i},\alpha_\text{i})$. It can be shown that
$\int_{-\infty}^0  \mathcal{T}_{y_\text{o}w_\text{o}}(\omega)d\omega\simeq\pi \gamma \rho({y_\text{o},w_\text{o}})$ for $\gamma$ smaller compared to the band width and gap, as elaborated in SI.~IV.~B,
where $\rho({y_\text{o},w_\text{o}})$ is the {mode density} of all occupied bands. Next, we introduce a weak gauge field in $x$-$z$ space (it can be implemented by controlling the tunneling phase~\cite{yuantwosynthetic} and examine the variation of $\rho({y_\text{o},w_\text{o}})$,
the layer resolved Chern number can be obtained according to the Streda formula~\cite{stredaformula}
$C(y_\text{o},w_\text{o})=2\pi\frac{\partial \rho({y_\text{o},w_\text{o}})}{\partial B_{xz}}$ (which can be generalized to higher dimensions), with $B_{xz}$ the corresponding weak magnetic field (see SI.~IV.~A for more details).
Shown in Figs.~\ref{mainsyn}(b) and (c) are our numerical simulations with mass term on only one corner and on all four corners, respectively.
The former leads to 1/2-quantized total Chern number mainly distributed around the corresponding corner, while the later lead to a total Chern number $C=2$ with each corner contributed 1/2.\\

\textbf{Conclusion and Discussion.} In conclusion, we have studied the realization, characterization and detection of general parity anomaly and Half-integer Hall response in high-dimensional higher-order topological insulators. Using the 4D second-order topological insulator as an example, we demonstrate that parity anomaly emerges at each corner hosting a single Dirac cone. We establish a general bulk-boundary correspondence by integrating nested Wilson loop theory with time-reversal polarization, leading to a set of 
$Z_2$ topological invariants that determine the number and momentum-space positions of Dirac cones at the corresponding corners.
These results can be extended to the general case of $d$-dimensional $n$-th order topological systems with time-reversal and reflection symmetric bulk, which can be constructed following the developed procedure.
Furthermore, we propose an experimental realization based on coupled cavity arrays, utilizing photonic internal degrees of freedom as synthetic dimensions. The parity anomaly induced half-integer Hall conductance can be measured using transmission spectra, which directly reflect the {mode density} and allow extracting layer-resolved Chern numbers via the Streda formula.
Our work advances the understanding of parity anomaly and higher-order topological phases, paves the way for exploring novel quantum phenomena and their applications in higher dimensional higher-order topological states.

Due to surface parity anomaly, the ordinary 3D TIs turn into axion insulators when the time-reversal-breaking boundary perturbations are introduced to induce half-quantized Hall current that leads to non-vanishing axion field in the bulk~\cite{photonicaxion,axionexp2018,axionexp2022}. Our results can be viewed as an higher-order extension of the axion insulators, and further studies on how axion field is distributed in such systems would be interesting. 
Furthermore, though we have focused on the higher-order parity anomaly on even-dimensional boundaries, our Hamiltonian construction procedure and topological characterization are applicable for higher-order chiral anomaly on odd-dimensional boundaries, where the mass term can induce the violation of chiral current conversation~\cite{parityanomalyorigin1,parityanomalyorigin2}. Simple schemes for experimentally probing of such chiral anomaly is highly desired. \\

\noindent{\bf \large Acknowledgments}\\
We thank Jin-Shi Xu, Ming Gong, Xiang-Fa Zhou, Mu Yang and Ze-Di Cheng for useful discussions. This work is supported by the Natural Science Foundation of China (Grants No. 12474366, and No. 11974334) and Innovation Program for Quantum Science and Technology (Grants No. 2021ZD0301200). X.-W. Luo also acknowledges support from the USTC start-up funding.




%

\begin{widetext}
\section*{Supplementary Information}

\onecolumngrid

\setcounter{figure}{0} \renewcommand{\thefigure}{S\arabic{figure}} %
\setcounter{equation}{0} \renewcommand{\theequation}{S\arabic{equation}}


\section{Corner states}
In Sec.~\ref{effectivecorner}, we provide some details on deriving the effective corner Hamiltonian. In Sec.~\ref{nonabelian berry phase}, we first describe the numerical method to calculate the Berry phase and Chern number of Hamiltonian Eq.~1 in the main text with open (periodic) boundary condition along $yw$-axis ($xz$-axis). To further prove the HIQHE at each corner, in Sec.~\ref{layerresolvedchern} we study how the Chern number is distributed over the $yw$-axis by calculating layer-resolved Chern number. Then in Sec.~\ref{differentcorner} we will discuss the effective corner state with open boundaries along different directions.

\subsection{Effective corner Hamiltonian}
\label{effectivecorner}
To derive the effective corner Hamiltonian for the model Eq.~1 in the main text, we consider a semi-infinite lattice along $y$ and $w$ directions. 
After the Fourier transformation $|k_x,k_z\rangle\otimes|y,w,\alpha\rangle=\frac{1}{\sqrt{N_yN_w}}\sum_{k_y,k_w}e^{i(k_yy+k_ww)}|k_x,k_z\rangle\otimes|k_y,k_w,\alpha\rangle$ (where $|\alpha\rangle$ denote the spin states of the Hamiltonian), we can transform the Hamiltonian Eq.~1 in main text to real space Hamiltonian:
\begin{equation}
    \begin{split}
        H_{\text{real}}(k_x,k_z)=&\sum_{y,w\geq0}\bigg\{\big[\sin k_x\Gamma_1+(m_1+\cos k_x)\Gamma_3+\sin k_z\Gamma_4+(m_2+\cos k_z)\Gamma_6\big]|y,w\rangle\langle y,w|
        \\&+\frac{|y,w,\rangle\langle y+1,w|-|y+1,w\rangle\langle y,w|}{2i}\Gamma_2+\frac{|y,w\rangle\langle y+1,w|+|y+1,w\rangle\langle y,w|}{2}\Gamma_3
        \\&+\frac{|y,w\rangle\langle y,w+1|-|y,w+1\rangle\langle y,w|}{2i}\Gamma_5+\frac{|y,w\rangle\langle y,w+1|+|y,w+1\rangle\langle y,w|}{2}\Gamma_6\bigg\}.
    \end{split}
    \label{hreal}
\end{equation}
Where $\Gamma_{(1)-(7)}=(\sigma_z\tau_zs_x,\sigma_y,\sigma_x,\sigma_z\tau_zs_y,\sigma_z\tau_y,\sigma_z\tau_x,\sigma_z\tau_zs_z)$, with $\sigma_i,\tau_j,s_k$ denoting the three different Pauli matrix. We only focus on the lower-left corner. According to Schrodinger equation $H_{\text{real}}(k_x,k_z)|\Psi_{\text{corner}}\rangle=E(k_x,k_z)|\Psi_{\text{corner}}\rangle$, using the ansartz $|\Psi_\text{corner}\rangle=\sum_{y,w\geq0}\mathcal{N}(\kappa_1)^y(\kappa_2)^w|y,w\rangle\otimes|\xi\rangle$, where $\mathcal{N}$ is a normalized factor. We solve the equation in four regions:
\\
\\
\textbf{1}. $y=0,w=0$ corner:
\begin{equation}
    \{\sin k_x\Gamma_1+(m_1+\cos k_x)\Gamma_3+\sin k_z\Gamma_4+(m_2+\cos k_z)\Gamma_6+\frac{\kappa_1}{2i}\Gamma_2+\frac{\kappa_1}{2}\Gamma_3+\frac{\kappa_2}{2i}\Gamma_5+\frac{\kappa_2}{2}\Gamma_6\}|\xi\rangle=E(k_x,k_z)|\xi\rangle
\end{equation}
\\
\textbf{2}. $y>0,w=0$ edge:
\begin{equation}
    \{\sin k_x\Gamma_1+(m_1+\cos k_x)\Gamma_3+\sin k_z\Gamma_4+(m_2+\cos k_z)\Gamma_6+\frac{\kappa_1-\kappa_1^{-1}}{2i}\Gamma_2+\frac{\kappa_1+\kappa_1^{-1}}{2}\Gamma_3+\frac{\kappa_2}{2i}\Gamma_5+\frac{\kappa_2}{2}\Gamma_6\}|\xi\rangle=E(k_x,k_z)|\xi\rangle
\end{equation}
\\
\textbf{3}. $y=0,w>0$ edge:
\begin{equation}
    \{\sin k_x\Gamma_1+(m_1+\cos k_x)\Gamma_3+\sin k_z\Gamma_4+(m_2+\cos k_z)\Gamma_6+\frac{\kappa_1}{2i}\Gamma_2+\frac{\kappa_1}{2}\Gamma_3+\frac{\kappa_2-\kappa_2^{-1}}{2i}\Gamma_5+\frac{\kappa_2+\kappa_2^{-1}}{2}\Gamma_6\}|\xi\rangle=E(k_x,k_z)|\xi\rangle
\end{equation}
\\
\textbf{4}. $y>0,w>0$ bulk:
\begin{equation}
    \begin{split}
    &\{\sin k_x\Gamma_1+(m_1+\cos k_x)\Gamma_3+\sin k_z\Gamma_4+(m_2+\cos k_z)\Gamma_6\\
    +&\frac{\kappa_1-\kappa_1^{-1}}{2i}\Gamma_2+\frac{\kappa_1+\kappa_1^{-1}}{2}\Gamma_3+\frac{\kappa_2-\kappa_2^{-1}}{2i}\Gamma_5+\frac{\kappa_2+\kappa_2^{-1}}{2}\Gamma_6\}|\xi\rangle=E(k_x,k_z)|\xi\rangle
    \end{split}
\end{equation}
Solving all the equations above, we find 
the corner state must satisfy
\begin{equation}
P_{y\leftarrow}|\xi\rangle=|\xi\rangle\rightarrow \sigma_z|\xi\rangle=+|\xi\rangle
\end{equation}
\begin{equation}
P_{w\downarrow}|\xi\rangle=|\xi\rangle\rightarrow \tau_z|\xi\rangle=+|\xi\rangle
\end{equation}
\begin{equation}
\kappa_1(k_x,k_z)=-m_1-\cos k_x
\end{equation}
\begin{equation}
\kappa_2(k_x,k_z)=-m_2-\cos k_z
\end{equation}
Where $P_{y\leftarrow}=\frac{I+i\Gamma_2\Gamma_3}{2}$ and $P_{w\downarrow}=\frac{I+i\Gamma_5\Gamma_6}{2}$ are projection operators determining the eigenstate of the spin $|\xi\rangle$, and we require $|\kappa_1(k_x,k_z)|<1$, $|\kappa_2(k_x,k_z)|<1$. 

We consider the region $0<m_1,m_2<2$, the above corner state solution only exists around the point of $k_x=k_z=\pi$, we can project the Hamiltonian to the corner-state subspace $\{|\xi\rangle\}=\{|\uparrow\uparrow\uparrow\rangle_{\sigma\tau s},|\uparrow\uparrow\downarrow\rangle_{\sigma\tau s}\}$ and
expand $k_x=\pi+\delta k_x,k_z=\pi+\delta k_z$, the effective Hamiltonian is reduced to Dirac-type 
\begin{equation}
    H_{\text{corner}}(k_x,k_z)=P_{y\leftarrow}P_{w\downarrow}HP_{w\downarrow}P_{y\leftarrow}=\sin k_xs_x+\sin k_ys_y=-\delta k_xs_x-\delta k_ys_y.
    \label{eq:corner_H_supp}
\end{equation}
Similarly, we can solve for the corner states for other three corners in $yw$-space. Each corner can host a single Dirac cone, leading to the higher-order parity anomaly.

To open a band gap of the corner Dirac cone and generate  half-integer quantized Hall effect (HIQHE), the key is to break both the parity symmetry and time-reversal symmetry by adding mass term $H_\text{M}$. The parity symmetry corresponds to reflection of one coordinate, e.g., $s_xH_{\text{corner}}(k_x,k_z)s_x=H_{\text{corner}}(k_x,-k_z)$.
Consider the corner Hamiltonian given above, we can prove that the mass term can only taking the forms of  $i\Gamma_1\Gamma_3$, $\Gamma_7$, $i\Gamma_2\Gamma_5\Gamma_7$, $i\Gamma_4\Gamma_6\Gamma_7$ or their linear combinations,
these terms break both parity symmetry and time-reversal symmetry $T=s_yK$. This is because,
the Clifford algebra satisfy:
\begin{equation}
\{\Gamma_j,\Gamma_j\}=2\delta_{jj'},
\Gamma_1\Gamma_2\Gamma_3\Gamma_4\Gamma_5\Gamma_6\Gamma_7=\pm i\mathcal{I}_8,
\end{equation}
and the mass term will open the band gap if and only if 
\begin{equation}
\{H_\text{M},s_x\}=0,\{H_\text{M},s_y\}=0
\label{dakainengdai}
\end{equation}
so we find four linearly independent operators $i\Gamma_1\Gamma_3=-s_z,\;\Gamma_7=\sigma_z\tau_zs_z,\;i\Gamma_2\Gamma_5\Gamma_7=\tau_zs_z,\;i\Gamma_4\Gamma_6\Gamma_7=\sigma_zs_z$ satisfying the above conditions.
As an example, we can consider the mass term: $H_\text{M}=M\cdot i\Gamma_1\Gamma_3$. It can be shown that $H_\text{M}$ breaks both parity symmetry and time-reversal symmetry $T=s_yK$. Such mass term will open a gap, leading to a massive Dirac cone and HIQHE, the direction of Hall conductance is determined by the sign of the effective Dirac mass.

\subsection{Numerical calculation of Berry phase}
\label{nonabelian berry phase}
In this subsection, we introduce numerical method to calculate Berry phase of our model. 
We consider open boundaries along $y$ and $w$ directions. We treat the crystal as a wide, pseudo-2D lattice by absorbing the $y,w$-component into the internal degrees of freedom of an enlarged unit cell that extends along the entire length of the crystal in the $x,z$ directions. The Hamiltonian and eigenvector can be described as follows:
\begin{equation}
	\begin{split}
		H={\left[\begin{array}{ccccccccccccccccccccccccc}
				H_{yy}&H_{y,y+1}&&&\\
                    H_{y,y-1}&H_{yy}&H_{y,y+1}&&\\
                    &H_{y,y-1}&H_{yy}&\cdot\cdot\cdot&\\
                    &&\cdot\cdot\cdot&\cdot\cdot\cdot&H_{y,y+1}\\
                    &&&H_{y,y-1}&H_{yy}\\
			\end{array}
			\right]}
	\end{split}
\label{twoopen1}
\end{equation}
where $H_{y,y+1}=I_{N_w}\otimes(\frac{\Gamma_2}{2i}+\frac{\Gamma_5}{2}),H_{y,y-1}=I_{N_w}\otimes(\frac{-\Gamma_2}{2i}+\frac{\Gamma_5}{2})$, and
\begin{equation}
	\begin{split}
		H_{yy}={\left[\begin{array}{ccccccccccccccccccccccccc}
				H_{yy}^{ww}&H_{yy}^{w,w+1}&&&\\
                    H_{yy}^{w,w-1}&H_{yy}^{ww}&H_{yy}^{w,w+1}&&\\
                    &H_{yy}^{w,w-1}&H_{yy}^{ww}&\cdot\cdot\cdot&\\
                    &&\cdot\cdot\cdot&\cdot\cdot\cdot&H_{yy}^{w,w+1}\\
                    &&&H_{yy}^{w,w-1}&H_{yy}^{ww}\\
			\end{array}
			\right]}
	\end{split}
\end{equation}
with $H_{yy}^{ww}=\sin k_x\Gamma_1+(m_1+\cos k_x)\Gamma_5+\sin k_z\Gamma_3+(m_2+\cos k_z)\Gamma_6+H_\text{M}$, and $H_{yy}^{w,w+1}=\frac{\Gamma_4}{2i}+\frac{\Gamma_6}{2}$, $H_{yy}^{w,w-1}=\frac{-\Gamma_4}{2i}+\frac{\Gamma_6}{2}$. The Hamiltonian of the system (Eq.~\ref{twoopen1}) is $(8N_yN_w)\times(8N_yN_w)$ matrix, and have $8N_yN_w$ eigenvectors that are orthogonal to each other. We consider the valence bands, it has $4N_yN_w$ valence bands, we label it $|u_{j,k_x,k_z}\rangle$, where $j=1,2,\cdot\cdot\cdot,4N_yN_w$

Following the method in~\cite{latticeberry}, we can calculate the Berry phase to get the total Chern number numerically. First
we compose all the valence wavefunction to a big matrix $|\varphi(k_x,k_z)\rangle$ defined as:
\begin{equation}
	\begin{split}
		|\varphi(k_x,k_z)\rangle={\left[\begin{array}{cccc}
		|u_{1,k_x,k_z}\rangle,&|u_{2,k_x,k_z}\rangle,&
            \cdot\cdot\cdot,&|u_{N_t,k_x,k_z}\rangle\\
			\end{array}
			\right]}
	\end{split}
\end{equation}
where $N_t=4N_yN_w$. We can discretize the $k_x,k_z$ space into $N_x,N_z$ segments with interval $\triangle k_x=\frac{2\pi}{N_x},\triangle k_z=\frac{2\pi}{N_z}$, and use the method in~\cite{latticeberry} to calculate the Berry phase, we define:
\begin{equation}
    \bar{M}_i(\vec{k})=|\varphi(\vec{k})\rangle\langle\varphi(\vec{k}+\triangle k_i\hat{e}_i)|,\;\bar{M}_{-i}(\vec{k})=|\varphi(\vec{k})\rangle\langle\varphi(\vec{k}-\triangle k_i\hat{e}_i)|
\end{equation}
where $\vec{k}=(k_x,k_z)$. The Berry phase is:
\begin{equation}
    \Omega_{k_xk_z}\triangle k_x\triangle k_z=\text{Im}\bigg\{\log\big[\frac{\bar{M}_x(k_x,k_z)\bar{M}_y(k_x+\triangle k_x,k_z)\bar{M}_{-x}(k_x+\triangle k_x,k_z+\triangle k_z)\bar{M}_{-y}(k_x,k_z+\triangle k_z)}{|\det[\bar{M}_x(k_x,k_z)\bar{M}_y(k_x+\triangle k_x,k_z)\bar{M}_{-x}(k_x+\triangle k_x,k_z+\triangle k_z)\bar{M}_{-y}(k_x,k_z+\triangle k_z)]|}\big]\bigg\}
\end{equation}
In numerically calculation, we set $N_x$ and $N_z$ big enough, for example, $N_x=N_z=36$ as we used in main text and supplementary material, the total Chern number reads
\begin{equation}
    C_{xz}=\frac{1}{2\pi}\int_0^{2\pi}\int_0^{2\pi}dk_xdk_z\text{Tr}[\Omega_{k_xk_z}]
\end{equation}
Notice that the above method can also be applied to the case with periodic boundary condition along $y$ and $w$. 
In the presence of mass term $H_\text{M}=M\cdot i\Gamma_1\Gamma_3$,
only when both $y$ and $w$ directions have open boundaries, 
the total Chern number is nonzero $C=\pm 2$ in the topological phases, as shown in Fig.~2 in the main text.
On the other hand, if $y$ or $w$ or both have periodic boundaries,
we have $C_{xz}=0$. 
So we can conclude that the nonzero Chern number origins from the four corner states.

\subsection{Layer-resolved Chern number}
\label{layerresolvedchern}
In this section, we introduce the layer-resolved Chern number which gives the distribution of Chern number in real space, and we find that the HIQHE originates from the corner state. Layer-resolved Chern number has been introduced to study surfaces of axion insulator, showing that the layer-resolved Chern number of top slice and bottom slice is close to half-integer quantized and of opposite sign~\cite{localmarker2,axionanglesupp}.

For the calculation of corner anomalous Hall conductivity, we implement local Chern marker~\cite{localmarker1} method to show that the Chern-Simons contribution to anomalous Hall conductivity can be expressed as a local and real space property. We can define a local Chern marker as
\begin{equation}
    C(\textbf{r})=-2\pi \text{Im}\langle \textbf{r}|P\textbf{r}Q\times Q\textbf{r}P|\textbf{r}\rangle
    \label{localchernshuzhi}
\end{equation}
Where $\textbf{r}$ is the position operator, and $|\textbf{r}\rangle$ denotes a state at spatial coordinate $\textbf{r}$, $P=\frac{1}{N_k}\sum_{m\mathbf{k}}|u_{m\mathbf{k}}\rangle_{\text{occ}}\langle u_{m\mathbf{k}}|$ and $Q=\frac{1}{N_k}\sum_{m\mathbf{k}}|u_{m\mathbf{k}}\rangle_{\text{unocc}}\langle u_{m\mathbf{k}}|$ are projection operator on the occupied (valence) and unoccupied (conduct) subspaces, respectively. The local Chern-Simons contribution to the anomalous Hall conductivity can be obtained via $\sigma_{\text{Hall}}(\textbf{r})=\frac{e^2}{h}C(\textbf{r})$, and the total Hall response is obtained by summing all the local contributions: $\sigma_{\text{Hall}}=\sum_\textbf{r}\sigma_{\text{Hall}}(\textbf{r})$.

We consider our model Eq.~1 in the main text, and set periodic (open) boundary condition along $x$-axis and $z$-axis ($y$-axis and $w$-axis). In this case, the local Chern marker (Eq.~\ref{localchernshuzhi}) can also be named as layer-resolved Chern number. After some manipulations similar as in~\cite{localmarker2}, we can get the expression of layer-resolved Chern number as:
\begin{equation}
    C_{xz}(y,w)=\frac{1}{2\pi}\int_{BZ}dk_xdk_z\text{Tr}[\Omega_{k_xk_z}\rho_\mathbf{k}(y,w)].
\end{equation}
Where $\Omega_{k_xk_z}$ is the Berry phase of the occupied bands that could be calculated using the method discussed in Sec.~\ref{nonabelian berry phase}, and local density of occupied bands is
\begin{equation}
    [\rho_{\mathbf{k}}]_{mm'}(y,w)=\sum_{\alpha}u_{m\mathbf{k}}^*(y,w,s)u_{m'\mathbf{k}}(y,w,\alpha)
\end{equation}
Since $\sum_{y,w}[\rho_{\mathbf{k}}(y,w)]_{mm'}=\delta_{mm'}$, so that by summing the layer-resolved Chern number of all the layers, we can get the total Chern number, the expression is: $C_{xz}=\sum_{y,w}C_{xz}(y,w)=\frac{1}{2\pi}\int_{BZ}dk_xdk_z\text{Tr}[\Omega_\mathbf{k}]$ as it should.

\subsection{Corner states with different open boundaries}
\label{differentcorner}
Previously, we focused on the corner states with open boundaries along $y$-axis and $w$-axis, leading to the corner Dirac cone and HIQHE in $x$-$z$ space. In this section, we discuss the corner states with different open boundaries of our model Eq.~1 in the main text. We rewrite the Hamiltonian here
\begin{eqnarray}
    H(\mathbf{k})&=&\sum_{i=1}^6h_i(\mathbf{k})\Gamma_i
    \label{Eqhamiltonian1supp}\\
    &=&\sin k_x\Gamma_1+\sin k_y\Gamma_2+\sin k_z\Gamma_3+\sin k_w\Gamma_4+(m_1+\cos k_x+\cos k_y)\Gamma_5+(m_2+\cos k_z+\cos k_w)\Gamma_6\nonumber
\end{eqnarray}%
with  $\Gamma_{(1)-(7)}=(\sigma_z\tau_zs_x,\sigma_y,\sigma_z\tau_zs_y,\sigma_z\tau_y,\sigma_x,\sigma_z\tau_x,\sigma_z\tau_zs_z)$.
Since the Hamiltonian has rotational symmetry $C_4^{x,y}=\frac{I-\Gamma_1\Gamma_2}{\sqrt{2}}$ and  $C_4^{z,w}=\frac{I-\Gamma_3\Gamma_4}{\sqrt{2}}$ with $C_4^{x,y}H(k_x,k_y,k_z,k_w)(C_4^{x,y})^{-1}=H(-k_y,k_x,k_z,k_w)$, so we can deduce that the parity anomaly and HIQHE could occur at the corner of $(q_1,q_2)$ with $q_1\in\{x,y\}$ and $q_2\in\{z,w\}$. Furthermore, the system also has a combined rotational symmetry when $m_1=m_2$: $C_4^{xy,zw}H(k_x,k_y,k_z,k_w)(C_4^{xy,zw})^{-1}=H(-k_z,-k_w,k_x,k_y)$, where $C_4^{xy,zw}=\frac{I-\Gamma_1\Gamma_3}{\sqrt{2}}\frac{I-\Gamma_2\Gamma_4}{\sqrt{2}}\frac{\Gamma_5+\Gamma_6}{\sqrt{2}}$, and similarly for $m_1=-m_2$: $\bar{C}_4^{xy,zw}H(k_x,k_y,k_z,k_w)[\bar{C}_4^{xy,zw}]^{-1}=H(-k_z+\pi,-k_w+\pi,k_x+\pi,k_y+\pi)$, where $\bar{C}_4^{xy,zw}=\frac{I+\Gamma_1\Gamma_3}{\sqrt{2}}\frac{I+\Gamma_2\Gamma_4}{\sqrt{2}}\frac{\Gamma_5-\Gamma_6}{\sqrt{2}}$. Along the parameter lines $m_1=\pm m_2$, our system corresponds to the intrinsic higher-order topological insulator where the bulk gap closing point coincides with the critical topological phase transition. Away from the symmetric lines, only edge spectrum closes its gap across the topological phase transition.

As we discussed above, topological corner states can appear for different open boundaries. With open boundaries along $(q_1,q_2)$  and periodic boundaries along the other two directions, where $q_1\in\{x,y\}$ and $q_2\in\{z,w\}$, parity anomaly will be realized at the corresponding corners.
For example, we consider open boundaries along $x$ and $z$ directions, periodic boundaries along $y$ and $w$ directions. The corner states solutions and effective corner Hamiltonian can be obtained 
following the method in Sec.~\ref{effectivecorner}. Alternatively, we can obtain the corner states based on previously solutions under proper $C_4$ rotations.
At the lower-right corner of $xz$ space, the spin states of the corner modes becomes  $|\xi\rangle\rightarrow C_4^{x,y}C_4^{z,w}|\xi\rangle$, and within this subspace, the effective Hamiltonian takes the same form as Eq.~(\ref{eq:corner_H_supp}).
The mass term can be obtained similarly 
$H_\text{M}\rightarrow C_4^{x,y}C_4^{z,w}H_\text{M}(C_4^{x,y}C_4^{z,w})^{-1}$, which takes the forms of $\Gamma_7$, $i\Gamma_2\Gamma_4$, $i\Gamma_1\Gamma_5\Gamma_7$, and $i\Gamma_3\Gamma_6\Gamma_7$ or their combinations.
The results for open (periodic) boundaries along $xw$ ($yz$) directions, as well as for open (periodic) boundaries along $yz$ ($xw$) directions can be calculated using the same method.
It is worthy to note that for open boundaries along $xy$ and periodic boundaries along $zw$ directions (or vice versa), no corner states exist.

As we mentioned in the main text, the mass term, and thereby the direction of HIQHE can be controlled independently for each corner. Also, the spin-state subspace of the corner state may be different from corner to corner, and thus the effective Dirac mass may have different signs for different corners, even we add the same mass term to them.  
In Fig.~\ref{figS2}, we plot the layer-resolved Chern number by adding different mass terms on four corners, the Hall conductance contributed by each corner is always half-integer quantized, though the sign may change for different mass terms.

\begin{figure}[htb]
    \centering
    \includegraphics[width=0.8\textwidth]{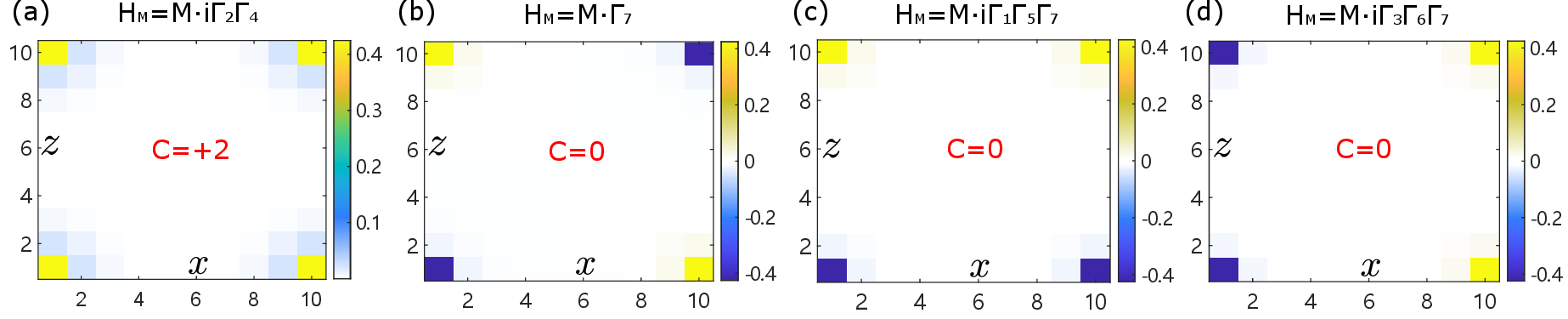}
    \caption{(a)-(d) Half-integer quantized Hall response of Hamiltonian (Eq.~\ref{Eqhamiltonian1supp}) with different mass terms. We consider open boundaries along $xz$-axis and periodic boundaries along $yw$-axis, and we add mass term $H_\text{M}$ only on four corner sites. Color bar shows the calculated layer-resolved Chern number which characterize the Chern-Simons contribution of each corner. By summing the layer-resolved Chern number over each corner region, we can get the corner Chern number is near 0.5 in all the cases above. Parameters are: $m_1=m_2=1.5,M=0.4,N_x=N_z=10,N_y=N_w=36$}
    \label{figS2}
\end{figure}

\begin{figure}
    \centering
    \includegraphics[width=0.8\textwidth]{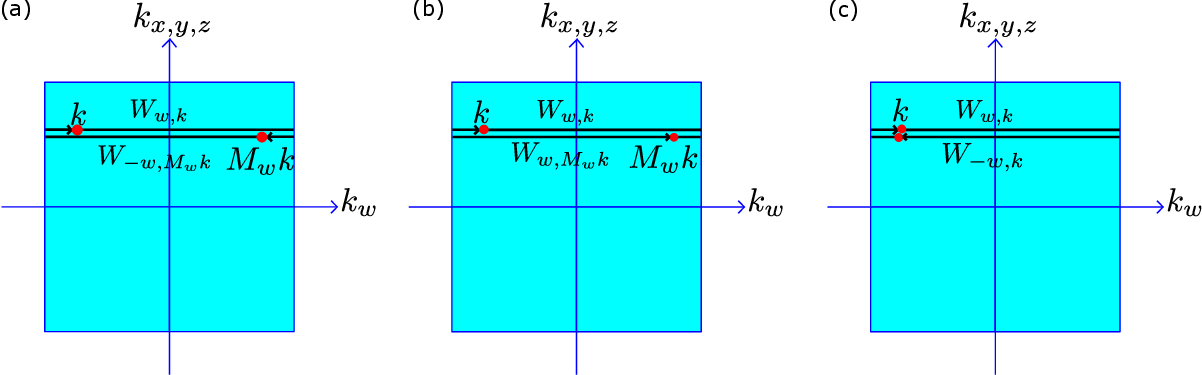}
    \caption{Relation between Wilson loop along $w$-axis at base point $\mathbf{k}$ with (a) reflection of both the direction and base point, (b)  reflection of only the base point, and (c) reflection of only the direction.}
    \label{wxkmirror}
\end{figure}

\section{Topological characterization}
\label{4d2ocharacterize}
In this section, we provide more details on the topological characterization and bulk-boundary correspondence of our model Eq.~1 in the main text.
We first calculate the Wilson loop along the $w$-axis, 
\begin{equation}
    W_{w,\mathbf{k}}=\mathcal{P}\exp[-i\int_{k_{w}}^{2\pi+k_{w}}A_w(k_x,k_y,k_z,k'_w)dk'_w]
\end{equation}
Where the base or start point $k_{w}$ does not affect the Wannier band, since $W_{w,\mathbf{k}}$ with different base point $k_{w}$ are related by a gauge (i.e., unitary) transformation $W_{w,k_x,k_y,k_z,k_w}=U W_{w,k_x,k_y,k_z,k'_w} U^\dag$. Such a gauge transformation does not affect our topological characterization and the following discussion. Without loss of generality,  we can set $k_{w}=0$ and omit the subscription in the Wilson loop operator.
The Wannier bands can be obtained by the eigen values of the Wilson loop operator
\begin{equation}
W_{w,\mathbf{k}}|\varepsilon_{w,j,\mathbf{k}}\rangle=e^{i2\pi\varepsilon_{w,j,\mathbf{k}}}|\varepsilon_{w,j,\mathbf{k}}\rangle.
\end{equation}
The Wannier-band basis can be obtained by project the Wilson loop operator to the occupied bands
\begin{equation}
    |\phi_{j,w}(\mathbf{k})\rangle=\sum_{m=1}^{N_\text{occ}}[|\varepsilon_{w,j,\mathbf{k}}\rangle]_m|u_{m,\mathbf{k}}\rangle.
\end{equation}
Numerically, the Wilson loop can be calculated using
$W_{w,\mathbf{k}}=\prod_{k_w} F_\mathbf{k}$. Where $[F_\mathbf{k}]_{mm'}=\langle u_{m,\mathbf{k}+\delta_{k_w}}|u_{m',\mathbf{k}}\rangle$, and $|u_{m,\mathbf{k}}\rangle$ is the eigenvector of Hamiltonian. In the following, we will first discuss the features of Wannier bands $\varepsilon_{w,j,\mathbf{k}}$ and Wannier wavefunction $|\varepsilon_{w,j,\mathbf{k}}\rangle$ under reflection and time-reversal symmetries. Then we provide some more details about the topological characterization.

\subsection{Wannier bands with reflection and time-reversal symmetry}
$\bullet$ First, we prove that the Wannier bands appear in $\pm \varepsilon$ pairs in the presence of $M_w$ reflection symmetry \begin{equation}
    M_wH_{\mathbf{k}}M_w^{\dagger}=H_{M_w\mathbf{k}}.
\end{equation}
From the definition of Wilson loop operator, we have
\begin{equation}
    W_{w,\mathbf{k}}\equiv W^{\dag}_{-w,\mathbf{k}}
\end{equation}
with $W_{-w,\mathbf{k}}$ the Wilson loop operator along $-w$ direction with base point $\mathbf{k}$ (see Fig.~\ref{wxkmirror} for the relation between different Wilson loops).
Due to reflection symmetry, we have
\begin{equation}
    M_w|u_{m',\mathbf{k}}\rangle=\sum_m |u_{m,M_w\mathbf{k}}\rangle B_{w,\mathbf{k}}^{mm'}
\end{equation} 
with
\begin{equation}
    B_{M_w,\mathbf{k}}^{mm'}=\langle u_{m,M_w\mathbf{k}}|M_w|u_{m',\mathbf{k}}\rangle
\end{equation}
a unitary matrix.
We note that $B_{w,\mathbf{k}}^{mm'}\neq0$ only if the eigen energies satisfy $E_{m,M_w\mathbf{k}}=E_{m',\mathbf{k}}$.
As a result, we find
\begin{equation}
    B_{w,\mathbf{k}}W_{w,\mathbf{k}}B_{w,\mathbf{k}}^{\dagger}=W_{-w,M_w\mathbf{k}}\equiv W_{w,M_w\mathbf{k}}^{\dagger}.
\end{equation}
Recall that the base point does not affect the Wannier band (eigenvalues of Wilson loop operator)
and the Wilson loop is a unitary operator, 
the eigenvalues must obey
\begin{equation}
    \{e^{i2\pi\varepsilon_{w,j,\mathbf{k}}}\}=\{e^{-i2\pi\varepsilon_{w,j,\mathbf{k}}}\}
\end{equation}
This can be seen more clearly by setting
$M_w\mathbf{k}=\mathbf{k}$, so we have
\begin{equation}
    B_{w,\mathbf{k}}W_{w,\mathbf{k}}B_{w,\mathbf{k}}^{\dagger}=W_{w,\mathbf{k}}^{\dagger}.
\end{equation}
For our case, the Wannier bands must satisfy $\varepsilon_{w,4,\mathbf{k}}=-\varepsilon_{w,1,\mathbf{k}},\varepsilon_{w,3,\mathbf{k}}=-\varepsilon_{w,2,\mathbf{k}}$.
Notice that the Wannier band is defined up to modular 1, so we can define $\varepsilon_{w,3,\mathbf{k}},\varepsilon_{w,4,\mathbf{k}}\in [0,1/2]$ as one Wannier sector and $\varepsilon_{w,1,\mathbf{k}},\varepsilon_{w,2,\mathbf{k}}\in [-1/2,0]$ as another Wannier sector.

$\bullet$ Second, we prove that in the presence of reflection symmetry $M_{xyz}=-iM_xM_yM_z$ and time-reversal symmetry $T^2=-1$, the two Wannier bands within each Wannier sector are degenerate, the corresponding Wannier-band basis are related by time-reversal symmetry $T$ and reflection symmetry $M_{xyz}$.
Due to the reflection symmetry (we use $M_{\bar{w}}$ to represent $M_{xyz}$ for short), we have
\begin{equation}
    M_{\bar{w}}|u_{m',\mathbf{k}}\rangle=\sum_m |u_{m,M_{\bar{w}}\mathbf{k}}\rangle B_{\bar{w},\mathbf{k}}^{mm'}
\end{equation} 
with $B_{\bar{w},\mathbf{k}}$ a unitary matrix and $B_{\bar{w},\mathbf{k}}^{mm'}\neq0$ only if the eigen energies satisfy $E_{m,M_{\bar{w}}\mathbf{k}}=E_{n,\mathbf{k}}$.
As a result, we find
\begin{equation}
    B_{\bar{w},\mathbf{k}}W_{w,\mathbf{k}}B_{\bar{w},\mathbf{k}}^{\dagger}=W_{w,M_{\bar{w}}\mathbf{k}}.
    \label{eq:M_Wilson_supp}
\end{equation}
Due to the time-reversal symmetry $T$, we have
\begin{equation}
    T|u_{m',\mathbf{k}}\rangle=\sum_m |u_{m,-\mathbf{k}}\rangle B_{T,\mathbf{k}}^{mm'}
\end{equation} 
with $B_{T,\mathbf{k}}$ a unitary matrix and $B_{T,\mathbf{k}}^{mm'}\neq0$ only if the eigen energies satisfy $E_{m,-\mathbf{k}}=E_{m',\mathbf{k}}$.
Using $\langle T a|T b\rangle=\langle b|a\rangle$, we find
\begin{equation}
    B_{T,\mathbf{k}}(W_{w,\mathbf{k}})^TB_{T,\mathbf{k}}^{\dagger}=W_{w,-\mathbf{k}}.
    \label{eq:T_Wilson_supp}
\end{equation}
$(W_{w,\mathbf{k}})^T$ is the transpose of $W_{w,\mathbf{k}}$.
Combining the symmetries $M_{\bar{w}}$ and $T$, we have
\begin{equation}
    B_{T,\mathbf{k}}^\dag B_{\bar{w},M_w\mathbf{k}} W_{w,M_w\mathbf{k}}B_{\bar{w},M_w\mathbf{k}}^\dag B_{T,\mathbf{k}}=(W_{w,\mathbf{k}})^T.
\end{equation}
That is
\begin{equation}
    B_{T\bar{w},\mathbf{k}}  W_{w,M_w\mathbf{k}} B_{T\bar{w},\mathbf{k}}^{\dagger}=(W_{w,\mathbf{k}})^T.
\end{equation}
Recall that the Wannier bands do not depend on the base point, so
we can choose $\mathbf{k}=M_w\mathbf{k}$, then both
$[B_{T\bar{w},\mathbf{k}} |\varepsilon_{w,j,\mathbf{k}}\rangle]^*$ and $|\varepsilon_{w,j,\mathbf{k}}\rangle$ are the eigenstates of $W_{w,\mathbf{k}}$ and they have the same eigenvalue. Notice that
the combined symmetry $TM_{\bar{w}}$ is anti-unitary satisfying $(TM_{\bar{w}})^2=-1$ [i.e., $(B_{T\bar{w},\mathbf{k}})^*B_{T\bar{w},\mathbf{k}}=-1$], and thus $[B_{T\bar{w},\mathbf{k}} |\varepsilon_{w,j,\mathbf{k}}\rangle]^*$ and $|\varepsilon_{w,j,\mathbf{k}}\rangle$ are distinct states, this means the Wannier bands must be two-fold degenerate, similar to the Kramer's degeneracy.

Now we deduce the symmetry properties of the Wannier band basis.
We have
\begin{equation}
    M_{\bar{w}}|\phi_{j,w}(\mathbf{k})\rangle=\sum_{m'=1}^{N_\text{occ}}[|\varepsilon_{w,j,\mathbf{k}}\rangle]_{m'}M_{\bar{w}}|u_{m',\mathbf{\mathbf{k}}}\rangle=\sum_{m}^{N_\text{occ}}\left\{\sum_{m'}^{N_\text{occ}} [|\varepsilon_{w,j,\mathbf{k}}\rangle]_{m'}B^{mm'}_{\bar{w},\mathbf{k}}\right\}|u_{m,M_{\bar{w}}\mathbf{k}}\rangle=\sum_{m}^{N_\text{occ}}[|\Tilde{\varepsilon}_{w,j,M_{\bar{w}}\mathbf{k}}\rangle]_m|u_{m,M_{\bar{w}}\mathbf{k}}\rangle.
\end{equation}
From Eq.~(\ref{eq:M_Wilson_supp}), we find
$|\Tilde{\varepsilon}_{w,j,M_{\bar{w}}\mathbf{k}}\rangle$ is an eigenstate of the Wilson loop $W_{w,M_{\bar{w}}\mathbf{k}}$ with the same eigenvalue as $|{\varepsilon}_{w,j,\mathbf{k}}\rangle$, where
\begin{equation}
[|\Tilde{\varepsilon}_{w,j,M_{\bar{w}}\mathbf{k}}\rangle]_m =   \sum_{m'}^{N_\text{occ}} [|\varepsilon_{w,j,\mathbf{k}}\rangle]_{m'}B^{mm'}_{\bar{w},\mathbf{k}}.
\end{equation}
As a result, the two Wannier-band states in each each Wannier sector are related by the reflection symmetry, that is
\begin{equation}
    \left\{M_{\bar{w}}|\phi_{j,w}(\mathbf{k})\rangle\right\}=   \left\{|\phi_{j,w}(M_{\bar{w}}\mathbf{k})\rangle\right\}.
\end{equation}
Similarly, based on Eq.~(\ref{eq:T_Wilson_supp}), each Wannier sector satisfies
\begin{equation}
    \left\{T|\phi_{j,w}(\mathbf{k})\rangle\right\}=   \left\{|\phi_{j,w}(-\mathbf{k})\rangle\right\}.
    \label{eq:T_Wannier_basis_supp}
\end{equation}
If we set the base point as the time-reversal invariant point $k_w=G_w$, then the two Wannier-band states in each each Wannier sector are related by the time-reversal symmetry.

$\bullet$ Third, we prove that, at the time-reversal invariant momenta, the Wannier-band basis must be the eigenstates of reflection symmetry $M_{\bar{w}}$, the corresponding eigenvalue are the same for the two Wannier-band states in each Wannier sector.  
We focus on the Wannier sector $\varepsilon_j\in [0,1/2]$ with $j=3,4$. Now let us consider the state at time-reversal invariant momenta
$G_x,G_y,G_z$ and the base point of the Wilson loop is chosen to be $k_w$.  Then we have
\begin{equation}
    \left\{M_{\bar{w}}|\phi_{3,w}(\mathbf{G}_{\bar{w}})\rangle,M_{\bar{w}}|\phi_{4,w}(\mathbf{G}_{\bar{w}})\rangle\right\}=   \left\{|\phi_{3,w}(\mathbf{G}_{\bar{w}})\rangle,|\phi_{4,w}(\mathbf{G}_{\bar{w}})\rangle\right\}
\end{equation}
with $\mathbf{G}_{\bar{w}}=(G_x,G_y,G_z,k_w)$.
Therefore, we can choose the Wannier-band basis to be the eigenstates of reflection symmetry (due to the degeneracy of the two Wannier bands) $M_{\bar{w}}|\phi_{j,w}(\mathbf{G}_{\bar{w}})\rangle=\eta_{j}(\mathbf{G}_{\bar{w}})|\phi_{j,w}(\mathbf{G}_{\bar{w}})\rangle$, since $M_{\bar{w}}^2=1$, we have $\eta_j=\pm1$. When both the energy and Wannier bands have gaps, $|\phi_{j,w}(\mathbf{G}_{\bar{w}})\rangle$ can always be chosen as a smooth function of the base point $k_w$, therefore, $\eta_{j}(\mathbf{G}_{\bar{w}})$ is independent of $k_w$. Without loss of generality, we consider $k_w=G_w$ and Eq.~\ref{eq:T_Wannier_basis_supp} becomes
\begin{equation}
    \left\{T|\phi_{3,w}(\mathbf{G})\rangle,T|\phi_{4,w}(\mathbf{G})\rangle\right\}=   \left\{|\phi_{3,w}(\mathbf{G})\rangle,|\phi_{4,w}(\mathbf{G})\rangle\right\}
\end{equation}
with $G=(G_x,G_y,G_z,G_w)$. Since time-reversal symmetry satisfy $T^2=-1$, thus we can set $T|\phi_{3,w}(\mathbf{G})\rangle=|\phi_{4,w}(\mathbf{G})\rangle$, and $T|\phi_{4,w}(\mathbf{G})\rangle=-|\phi_{3,w}(\mathbf{G})\rangle$. Then 
$T^2|\phi_{3,w}(\mathbf{G})\rangle=TM_{\bar{w}}TM_{\bar{w}}|\phi_{3,w}(\mathbf{G})\rangle=-\eta_4\eta_3|\phi_{3,w}(\mathbf{G})\rangle$, and thus we must have $\eta_3=\eta_4$.

\subsection{Topological characterization}
The model Eq.~1 in the main text posseses reflection $M_x,\;M_y,\;M_z,\;M_w$ and time-reversal $T$ symmetries with  $T^2=-1$. Based on the discussion above, the Wannier band structures satisfy: $\varepsilon_{w,1,\mathbf{k}}=\varepsilon_{w,2,\mathbf{k}}=-\varepsilon_{w,3,\mathbf{k}}=-\varepsilon_{w,4,\mathbf{k}}$. Similarly for the $y$-direction Wannier bands: $\varepsilon_{y,1,\mathbf{k}}=\varepsilon_{y,2,\mathbf{k}}=-\varepsilon_{y,3,\mathbf{k}}=-\varepsilon_{y,4,\mathbf{k}}$. 
The Wannier bands are related to the boundary properties, and we consider only the top or bottom boundary along $w$ by focusing on the Wannier sector: $\varepsilon_j\in(0,\frac{1}{2})$.
According to the properties of the Wannier-band basis,
we can generalize the $Z_2$ characterization of topological insulators in Ref.~\cite{3dclassification3d1,3dclassification3d2,3dclassification3d3,yuruinonabelian}, we establish the bulk-boundary correspondence by calculating the ``nested" time-reversal spin polarization using the Wannier-band basis. 
We consider time-reversal invariant momenta and use the method in \cite{3dclassification3d1,3dclassification3d2,3dclassification3d3} to define  the nested time-reversal polarization $\nu_y^{\varepsilon_w}(G_x,G_z)$  along $y$ through
\begin{equation}
    (-1)^{\nu_y^{\varepsilon_w}(G_x,G_z)}=\prod_{G_y=0,\pi}\eta^{\varepsilon_w}(\textbf{G})
\end{equation}
where $\eta^{\varepsilon_w}(\textbf{G})$ is the eigenvalue (parity) of the reflection symmetry $M_{xyz}|\phi_{3,w}(\textbf{G})\rangle=\eta^{\varepsilon_w}(\textbf{G})|\phi_{3,w}(\textbf{G})\rangle$ with eigenvalue $\eta^{\varepsilon_w}(\textbf{G})=\pm1$.
Notice that $\eta^{\varepsilon_w}(\mathbf{G})$ is associated to the 
Pfaffian $\eta^{\varepsilon_w}(\mathbf{G}) ={\text{Pf}[S^{\varepsilon_w}(\mathbf{G})]}/{\sqrt{\text{Det}[S^{\varepsilon_w}(\mathbf{G})]}}$ with skew matrix $S^{\varepsilon_w}_{j'j}(\mathbf{G})\equiv\langle\phi_{j',w}(\mathbf{G})|T|\phi_{j,w}(\mathbf{G})\rangle$ under transverse gauge choice~\cite{3dclassification3d3}.
Notice that the Wannier-band basis satisfy Eq.~(\ref{eq:T_Wannier_basis_supp}), so the matrix $S^{\varepsilon_w}(\mathbf{G})$ is anti-symmetric and the Pfaffian is well defined.

Similarly, we can calculate the Wilson loop along the $y$ and define the nested time-reversal spin polarization $\nu_w^{\varepsilon_y}(G_x,G_z)$ along $w$ direction through
\begin{equation}
    (-1)^{\nu_w^{\varepsilon_y}(G_xG_z)}=\prod_{G_w=0,\pi}\eta^{\varepsilon_y}(\textbf{G})
\end{equation}
We can extend the $Z_2\times Z_2$ characterization of 2D second-order topological insulator~\cite{bbh111,bbh222} and define our topological invariants as:
\begin{equation}
    \nu_{yw}^{G_x,G_z}=\nu_y^{\varepsilon_w}(G_x,G_z)\nu_w^{\varepsilon_y}(G_x,G_z).
\end{equation}
There are four topological invariant depending
on the values of $(G_x,G_z)$.
When $\nu_{yw}^{G_x,G_z}=1$, we can deduce that effective massless Dirac cone emerges near the real-space corner in $y,w$-direction and located at momentum $(G_x,G_z)$ in the $k_x,k_z$-space. We can define $\nu_{yw}^{\text{total}}=(\nu_{yw}^{00}+\nu_{yw}^{0\pi}+\nu_{yw}^{\pi0}+\nu_{yw}^{\pi\pi}) \text{ mod 2}$ to characterize the odd or even property of the total number of corner Dirac cones at each corner, and higher-order topological parity anomaly is realized when $\nu_{yw}^{\text{total}}$ is odd. 
We can redefine the four topological invariants as:
\begin{equation}
    (\nu_{yw}^{\text{total}};\nu_{yw}^{00},\nu_{yw}^{0\pi},\nu_{yw}^{\pi\pi})
\end{equation}
with $\nu_{yw}^{00},\;\nu_{yw}^{0\pi},\;\nu_{yw}^{\pi\pi}$ characterizing the momentum-space position of corner Dirac cone. 

The topological invariants of model Eq.~(1) in the main text is: (1;001) for the topological phase $T_1$ with $0<m_1,m_2<2$; (0;000) for the trivial phase $N$. In topological phase $T_1$, one effective Dirac cone emerges near each corner of $y,w$-space and located at $(G_x,G_z)=(\pi,\pi)$ in $k_x,k_z$ space. 
We want to point out that, our bulk-boundary correspondence theory can be applied to general higher-order topological parity anomalies, as we will discuss in the following (see Sec.~\ref{anothermodelii}).

\section{Generalization to higher-dimensional higher-order cases}
In this section, we provide more details on the general method to construct and characterize $d$-dimensional $n$-order topological insulators supporting effective corner Dirac cones, we also discuss the corresponding Hall-type response of the corner state. The general construction method will be discussed in Sec.~\ref{dncase}. 
Then we give the general Hall-type response of the corner states and their relation to the layer-resolved even-dimensional Chern number in Sec.~\ref{layer2nd}.
We further discuss two additional examples:
Another 4D second-order topological model 
in Sec.~\ref{anothermodelii} and a 6D second-order topological insulator model in Sec.~\ref{6d2omodelchap}, both support corner Dirac cones. 

\subsection{General model construction method and the bulk-boundary correspondence}
\label{dncase}
We give a general method to construct $d$-dimensional $n$-order topological insulator supporting corner Dirac cone. We divide $X=\{x_1,\cdot\cdot\cdot,x_d\}$ into $n$ set: $X=\{X_1,\cdot\cdot\cdot,X_n\}$, each set $X_j$ contains $d_j>0$ dimensions with $\sum_jd_j=d$. The Hamiltonian of the system is:
\begin{equation}
    H(\mathbf{k})=\sum_{j=1}^{d}\sin k_j\Gamma_j+\sum_{j=1}^n[m_j+\sum_{l\in X_j}\cos k_j]\Gamma_{j+d}
    \label{eq:general_Ham_supp}
\end{equation}
We construct $\Gamma$ matrix using $p_\text{corner}=\lceil\frac{d+n-1}{2}\rceil$ Paulis (where $\lceil\;\rceil$ denotes round up to nearest integer). In particular, we have 
$\Gamma^{(p)}=\{\sigma_x^{(p)},\sigma_y^{(p)},\sigma_z^{(p)}\otimes\Gamma^{(p-1)}\}$ with
$\Gamma^{(1)}=\{\sigma_x^{(1)},\sigma_y^{(1)},\sigma_z^{(1)}\}$,
so $\Gamma^{(p)}=\{\Gamma_1,\cdots,\Gamma_{d+n},\cdots\}$ contains at least $d+n$ elements, similar to that in Ref.~\cite{dimensionalreduction}.

We consider open boundaries along $(q_1,\cdot\cdot\cdot,q_n)$ with $q_j\in X_j$, and periodic boundaries along other direction $(q_{n+1},\cdot\cdot\cdot,q_d)$. For convenience, we only consider the corner states near the corner $(q_1q_2\cdot\cdot\cdot q_n)=(00\cdot\cdot\cdot0)$. Using the ansartz: $|\Psi_{\text{corner}}\rangle=\mathcal{N}\prod_{j=1}^{n}(\kappa_j)^{q_j}|q_1q_2\cdot\cdot\cdot q_n\rangle\otimes|\xi\rangle$, we can solve for the corner state, the spin component is determined by a set of projection operators $\{P_{q_j,0}\}$ (where $j=1,2,\cdot\cdot\cdot,n,q_j\in X_j$) with
\begin{equation}
    P_{q_j,0}=\frac{I+i\Gamma_j\Gamma_{j+d}}{2}.
\end{equation}
We have
\begin{equation}
    P_{q_j,0}|\xi\rangle=+|\xi\rangle
\end{equation}
\begin{equation}
    m_j+\kappa_j+\sum_{l\neq j,l\in X_j}\cos k_l=0
\end{equation}
The existence of corner states requires that $|\kappa_j|<1$, which can be satisfied if and only if $|m_j|<d_j$.

Now we discuss the effective Hamiltonian of the Dirac-like massless corner states. The effective Hamiltonian of the corner states is:
\begin{equation}
    H_{\text{corner}}=(P_{q_1,0}P_{q_2,0}\cdots P_{q_n,0})H(\mathbf{k})(P_{q_n,0}\cdots P_{q_2,0}P_{q_1,0})
\end{equation}
The Dirac-like massless corner modes lie near the region $\sin k_{q_j}^0=0,\;\cos k_{q_j}^0=\varsigma({k_{q_j}})=\pm1$ (with $j=n+1,\cdots,d$). By writing $k_{q_j}=k_{q_j}^0+\delta k_{q_j}$, we can find effective Hamiltonian of the system:
\begin{equation}
    H_{\text{corner}}=\sum_{j=n+1}^d\varsigma(k_{q_j})\delta k_{q_j}\Gamma'_{q_j},
\end{equation}
where
\begin{equation}
    \Gamma'_{q_j}=(P_{q_1,0}P_{q_2,0}\cdot\cdot\cdot P_{q_n,0})\Gamma_i(P_{q_n,0}\cdot\cdot\cdot P_{q_2,0}P_{q_1,0}), \text{ with } j=d+1,\cdots,d.
    \label{gammapie}
\end{equation}
We see that the corner states is represented by
$p_\text{corner}=\lceil\frac{d-n-1}{2}\rceil$ Paulis, leading to $2^{p_\text{corner}}$-fold degeneracy.
Note that $k_{q_j}^0$ should be proper chosen to make $|\kappa_j|<1$ satisfied.
Therefore, the system can support $(d-n)$-dimensional 
massless Dirac-like modes.
When $(d-n)$ is even (odd), the system realizes a general higher-order topological parity (chiral) anomaly. For the parity anomaly, proper mass term $H_\text{M}$ would open the corner gaps, leading to anomalous Hall conductivity characterized by $(d-n)/2$-th Chern number. We can use the method discussed in Sec.~\ref{layer2nd} to calculate layer-resolved $(d-n)$-dimensional $(d-n)/2$-th Chern number which reflect the Chern-Simons contribution to the Hall conductance of the corner state.

To extend our bulk-boundary correspondence to the general case of higher-order topological parity anomaly, we can define the nested time-reversal polarization $\nu_{q_j}^{\varepsilon_{\bar{q}_j}}$ along $q_j$ using the Wannier-band basis after evaluating the nested Wilson loops along $\bar{q}_j=(q_1,\cdots,q_{j-1},q_{j+1},q_n)$, which requires calculating the Wilson loop operator along $n-1$ dimensions, and the order of the dimensions in the calculation does not affect the final results. 
The $Z_2$ topological invariants then read $\nu_{q_1\cdots q_n}=\prod_j \nu_{q_j}^{\varepsilon_{\bar{q}_j}}$. We have $2^{d-n}$ 
$Z_2$ invariants associated with the symmetric momenta in the $d-n$ dimensions. 
In fact, our topological characterization relies solely on time-reversal and reflection symmetries (with $C_4$ symmetry non-essential), making it applicable to any models preserving these symmetries, including 3D second-order hinge states and higher-order chiral anomaly with odd $(d-n)$.

\subsection{General layer-resolved Chern number}
\label{layer2nd}
When $d-n$ is even, by adding proper mass term $H_\text{M}$ to open the gap of the corner states, the general Hall response of each Dirac cone can be written as (with Einstein summation notation) \cite{dimensionalreduction,localsecondchern}
\begin{equation}
    j^{i_{1}}(q_1,q_2,\cdot\cdot\cdot,q_n)=\frac{C_{\frac{d-n}{2}}(q_1,q_2,\cdot\cdot\cdot,q_n)}{2^{\frac{d-n}{2}-1}\cdot(2\pi)^{\frac{d-n}{2}}}\epsilon_{i_{1}i_{2}\cdot\cdot\cdot i_{d-n}}B_{i_{2}i_{3}}\cdot\cdot\cdot B_{i_{d-n-2}i_{d-n-1}}E_{i_{d-n}}
    \label{jlocalhigh}
\end{equation}
Where $B_{ij}$ and $E_{j}$ are gauge magnetic field and gauge electric field, and $(i_1,i_2,\cdot\cdot\cdot,i_{d-n})\in(q_{n+1},q_{n+2},\cdot\cdot\cdot,q_{d})$. 
$C_{\frac{d-n}{2}}(q_{1},q_{2},\cdot\cdot\cdot ,q_{n})$ is the layer-resolved $(d-n)/2$-th Chern number, which reads (with Einstein summation notation)
\begin{equation}
\begin{split}
    C_{\frac{d-n}{2}}(q_{1},q_{2},\cdot\cdot\cdot,q_n)&=\frac{1}{(\frac{d-n}{2})!2^{(d-n)/2}(2\pi)^{(d-n)/2}}\int d^{d-n}\mathbf{k}\epsilon^{i_1i_2\cdot\cdot\cdot i_{d-n}}\text{Tr}[\Omega_{i_1i_2}\Omega_{i_3i_4}\cdot\cdot\cdot\Omega_{i_{d-n-1}i_{d-n}}]\rho_k(q_{1},q_{2},\cdot\cdot\cdot,q_n)\label{locallocalhigh}
\end{split}
\end{equation}
with occupied-band density
\begin{equation}
    [\rho_{\mathbf{k}}]_{mm'}=\sum_\alpha u_{m\mathbf{k}}^*(q_{1},q_{2},\cdot\cdot\cdot,q_n,\alpha)u_{m'\mathbf{k}}(q_{1},q_{2},\cdot\cdot\cdot,q_n,\alpha)
\end{equation}
Where $\textbf{k}=(q_{n+1},q_{n+2},\cdot\cdot\cdot,q_d)$, and $u_{m\textbf{k}}(q_1,q_2,\cdot\cdot\cdot,q_n,\alpha)$ denotes the eigen-state component of $(q_1,q_2,\cdot\cdot\cdot,q_n,\alpha)$ site (spin) for $m$-th occupied energy band.  We consider periodic (open) boundary condition along $q_{n+1}q_{n+2}\cdot\cdot\cdot q_{d}$ ($q_{1}q_{2}\cdot\cdot\cdot q_n$) directions.
Layer means the site positions in the space $q_1,q_2,\cdot\cdot\cdot, q_n$, while the $(d-n)/2$-th Chern number capture the electromagnetic response in the 
$(d-n)$-dimensional space $q_{n+1},\cdots,q_d$.

For example, the layer-resolved first Chern calculated in the main text is $ C_1(\textbf{r})=2\pi i\epsilon_{jj'}\langle \textbf{r}|Pr_jQr_{j'}P|\textbf{r}\rangle$.
The layer-resolved second Chern number~\cite{localsecondchern} corresponding to $d-n=4$ reads (with Einstein summation notation)
\begin{equation}
    C_2(\textbf{r})=-2\pi^2\epsilon_{i_1i_2i_3i_4}\langle \textbf{r}|Pr_{i_1}Qr_{i_2}Pr_{i_3}Qr_{i_4}P|\textbf{r}\rangle
\end{equation}
Where $(i_1,i_2,i_3,i_4)\in(q_{n+1},q_{n+2},q_{n+3},q_{n+4})$. After some manipulations, we can get the result of layer-resolved second Chern number as (with Einstein summation notation)
\begin{equation}
    C_2(q_1,\cdots,q_n)=\frac{1}{32\pi^2}\int d^4\mathbf{k}\epsilon_{i_1i_2i_3i_4}\text{Tr}[\Omega_{i_1i_2}\Omega_{i_3i_4}\rho_\mathbf{k}(q_1,\cdots,q_n)] 
\end{equation}
with
\begin{equation}
    [\rho_{\mathbf{k}}]_{mm'}=\sum_\alpha u_{m\mathbf{k}}^*(q_1,\cdots,q_n,\alpha)u_{m'\mathbf{k}}(q_1,\cdots,q_n,\alpha)
\end{equation}
Where $\textbf{k}=(k_{q_{n+1}},\cdots,k_{q_{d}})$. Since $\sum_{q_1,\cdots,q_n}[\rho_{\mathbf{k}}]_{mm'}(q_1,\cdots,q_n)=\delta_{mm'}$, so that by summing layer-resolved second Chern number of all the layers, we can get the total second Chern number, the expression is: $C_2=\sum_{q_1,\cdots,q_n}C_2(q_1,\cdots,q_n)=\frac{1}{32\pi^2}\int d^4\mathbf{k}\epsilon_{i_1i_2i_3i_4}\text{Tr}[\Omega_{i_1i_2}\Omega_{i_3i_4}]$ which characterized the electromagnetic response in the space $q_{n+1},\cdots,q_{n+4}$.

\subsection{Model II: 4-dimensional second-order Topological insulator}
\label{anothermodelii}
In this section, we discuss another 4-dimensional second-order topological insulator model that also realizes parity anomaly at the corners. From the general Hamiltonian Eq.~\ref{eq:general_Ham_supp}, we can construct the Hamiltonian of a different 4D second order topological insulator model (we denote it as model II):
\begin{equation}
    \begin{split}
        H(\mathbf{k})&=\sum_{i=1}^6h_i(\mathbf{k})\Gamma_i\\
        &=\sin k_x\Gamma_1+\sin k_y\Gamma_2+\sin k_z\Gamma_3+\sin k_w\Gamma_4+(m_1+\cos k_x+\cos k_y+\cos k_z)\Gamma_5+(m_2+\cos k_w)\Gamma_6
    \end{split}
    \label{1add3model}
\end{equation}
Where $\Gamma_{(1)-(7)}=(\sigma_z\tau_zs_x,\sigma_y,\sigma_z\tau_zs_y,\sigma_z\tau_y,\sigma_x,\sigma_z\tau_x,\sigma_z\tau_zs_z)$ is the same as that in main text. The energy of this Hamiltonian is $E(\text{k})=\pm\sqrt{\sum_{i=1}^6h_i(\text{k})^2}$. The energy gap closes at eight points with $|m_1|=1,3$ and $|m_2|=1$ in the $m_1$-$m_2$ parameter space. This Bloch Hamiltonian has time-reversal symmetry, reflection symmetry same as that in main text: $T=s_yK$, $M_x=s_y$, $M_y=\sigma_x\tau_zs_z$, $M_z=s_x,M_w=\tau_xs_z$, but its $C_4$ rotational symmetry is different from the model in main text, here the symmetries of model II are: $C_4^{x,y}=\frac{I-\Gamma_1\Gamma_2}{\sqrt{2}}=\exp(i\frac{\pi}{4}\sigma_x\tau_zs_x)$, $C_4^{x,z}=\frac{I-\Gamma_1\Gamma_3}{\sqrt{2}}=\exp(-i\frac{\pi}{4}s_z)$, $C_4^{y,z}=\frac{I-\Gamma_2\Gamma_3}{\sqrt{2}}=\exp(-i\frac{\pi}{4}\sigma_x\tau_zs_y)$.
The band structures of model II are shown in Figs.~\ref{add13modelsupp} (a)-(b)

\begin{figure}[htb]
    \centering
    \includegraphics[width=0.9\textwidth]{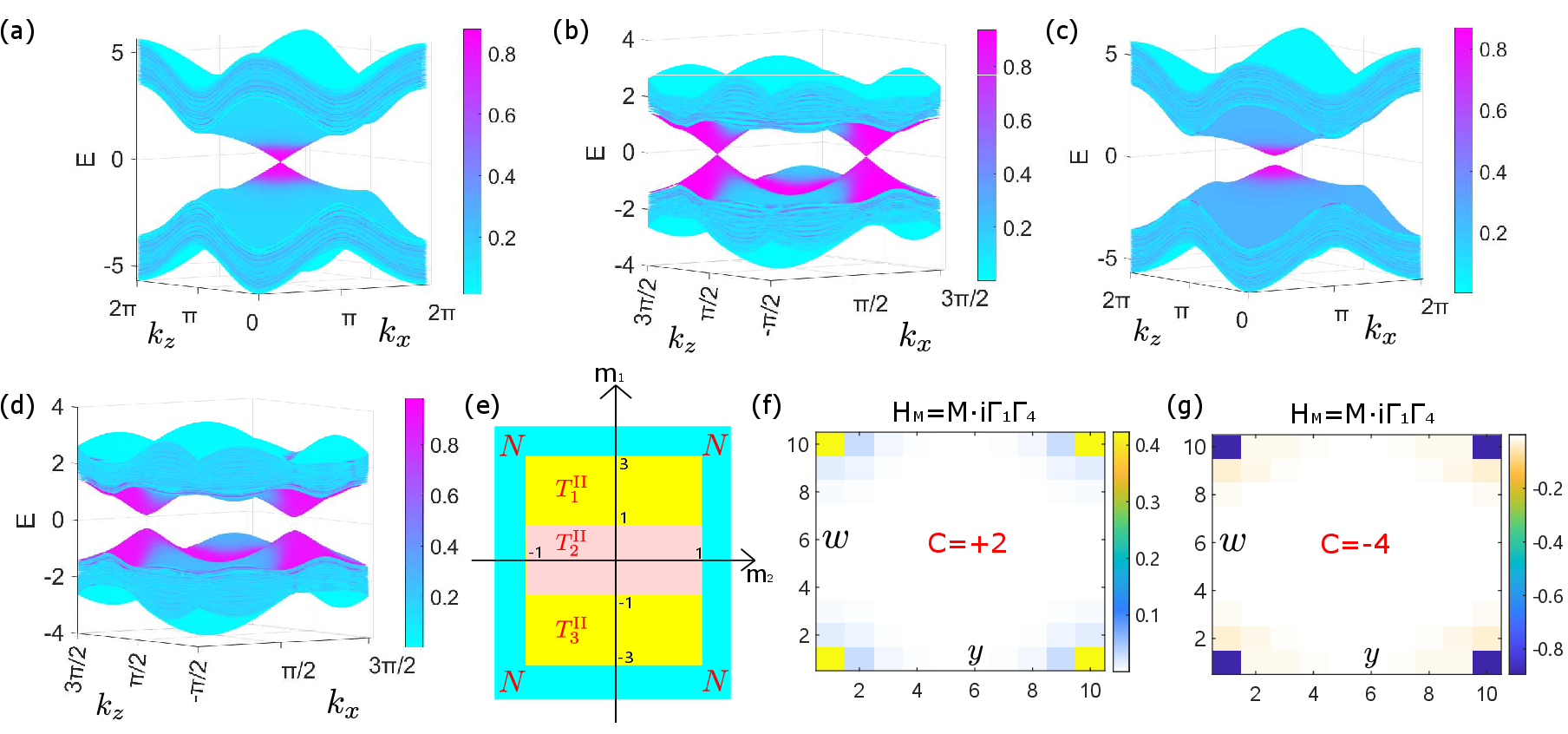}
\caption{[(a) and (b)] Energy bands of model II Eq.~\ref{1add3model} with $m_1=2.5$ in (a), and $m_1=0.5$ in (b).  The system hosts one effective Dirac cone in (a), two effective Dirac cones in (b). [(c) and (d)] Energy bands with mass terms $H_\text{M}=M\cdot i\Gamma_1\Gamma_4$ on four corner sites, $m_1=2.5$ in (c), and $m_1=0.5$ in (d). The mass terms break parity and time-reversal symmetries and open a band gap. The color bar in (a)-(d) reflects the population of the wave function on the four corner sites. 
(e) Phase diagram in $m_1$-$m_2$ parameter space. [(f) and (g)] Layer-resolved Chern number of model in the presence of mass term $H_\text{M}=M\cdot i\Gamma_1\Gamma_4$ on the four corner sites, 
with $m_1=2.5$ in (f) and $m_1=0.5$ in (g). Each corner contributes 1/2-quantized Hall effect in (f) and 1-quantized Hall effect in (g). Other parameters are: $N_x=N_z=36,N_y=N_w=10,M=0.4,m_2=0.5$.}
    \label{add13modelsupp}
\end{figure}

We consider open (periodic) boundary conditions along
$y$-$w$ ($x$-$z$) directions, discuss the parity anomaly near the corner at $y=w=0$. Following the method in Sec.~\ref{effectivecorner}, we can solve for the projection operator as $P_{y\leftarrow}=\frac{I+i\Gamma_2\Gamma_5}{2}=\frac{I+\sigma_z}{2},\;P_{w\downarrow}=\frac{I+i\Gamma_4\Gamma_6}{2}=\frac{I+\tau_z}{2}$, the ansartz $|\Psi_{\text{corner}}\rangle=\sum_{y,w\geq0}\mathcal{N}(\kappa_1)^y(\kappa_2)^w|y,w\rangle\otimes|\xi\rangle$, where $\mathcal{N}$ is normalization factor, $|\xi\rangle$ is the eigenstate of the projector with $+1$ eigenvalue, and $\kappa_1=-m_1-\cos k_x-\cos k_z,\kappa_2=-m_2$. Besides the trivial phase, we have three topological phases [as shown in Fig.~\ref{add13modelsupp} (e)]:
\begin{itemize}
    \item Topological phase $T_1^{\text{II}}$ for $1<m_1<3$ and $|m_2|<1$, one effective Dirac cone lies near $(k_x,k_z)=(\pi,\pi)$, we set $k_x=\pi+\delta k_x,k_z=\pi+\delta k_z$, and obtain the effective Hamiltonian as $H_\text{corner}=-\delta k_xs_x-\delta k_zs_y$
    \item Topological phase $T_2^{\text{II}}$ for $-1<m_1<1$ and $|m_2|<1$, two effective Dirac cone lie near $(k_x,k_z)=(0,\pi)$ and $(k_x,k_z)=(\pi,0)$, the effective Hamiltonian is $H_{\text{corner}}=\pm k_xs_x\mp k_zs_y$, the two Dirac cone has the same chirality.
    \item Topological phase $T_3^{\text{II}}$ for $-3<m_1<-1$ and $|m_2|<1$, one effective Dirac cone lies near $(k_x,k_z)=(0,0)$, the effective Hamiltonian is $H_{\text{corner}}=\delta k_xs_x+\delta k_zs_y$.
\end{itemize}

Similar as that discussed in Sec.~\ref{effectivecorner}, corner gap can be opened by mass term $H_\text{M}$ taking the forms of $M\cdot i\Gamma_1\Gamma_3,M\Gamma_7,M\cdot i\Gamma_2\Gamma_5\Gamma_7,M\cdot i\Gamma_4\Gamma_6\Gamma_7$ or there combinations. For example, we can add the mass term $H_\text{M}=M\cdot i\Gamma_1\Gamma_3$ on the four corner sites, leading to gaped corner states, as shown in Figs.~\ref{add13modelsupp} (c)(d).
In the presence of mass terms
$H_\text{M}=M\cdot i\Gamma_1\Gamma_3$ on the four corners, 
we obtain 1/2-quantized (1-quantized) Hall effect at each corner in phases
$T^{\text{II}}_1$ and $T^{\text{II}}_3$ (phase $T^{\text{II}}_2$), since there is one (two)
Dirac cone(s) at each corner.
In Figs.~\ref{add13modelsupp} (f) and (g), we plot the layer-resolved Chern number which reflects the Hall response of the corner states.
Finally, we discuss the cases with open boundary conditions along different directions. For open (periodic) boundary conditions along $x,w$  ($y,z$) directions,  as well as open (periodic) boundary conditions along $z,w$  ($x,y$) directions, the corner physics are similar as that for open (periodic) boundary conditions along $y,w$  ($x,z$) directions. However,
for open boundary conditions along $x,y$ or $x,z$ or $y,z$, the system does not support corner modes.

To verify the bulk-boundary correspondence, we calculated the Wannier bands and topological invariants, as shown in Fig.~\ref{figsuppcharacterize2}. 
We find that: $(\nu_\text{total};\nu_{0,0},\nu_{0,\pi},\nu_{\pi,\pi})$ is (1;001) in the $T^{\text{II}}_1$ phase with a single Dirac cone around $(G_x,G_z)=(\pi,\pi)$
at each corner;
$(\nu_\text{total};\nu_{0,0},\nu_{0,\pi},\nu_{\pi,\pi})$ is (0,010)  in the $T^{\text{II}}_2$ phase with two Dirac cones around $(G_x,G_z)=(0,\pi),(\pi,0)$
at each corner; $(\nu_\text{total};\nu_{0,0},\nu_{0,\pi},\nu_{\pi,\pi})$ is (1;100)  in the $T^{\text{II}}_3$ phase with a single Dirac cone around $(G_x,G_z)=(0,0)$
at each corner.
Higher order topological parity anomaly is realized in phases $T^{\text{II}}_1$ and $T^{\text{II}}_3$. 

\begin{figure}
    \centering
    \includegraphics[width=0.98\textwidth]{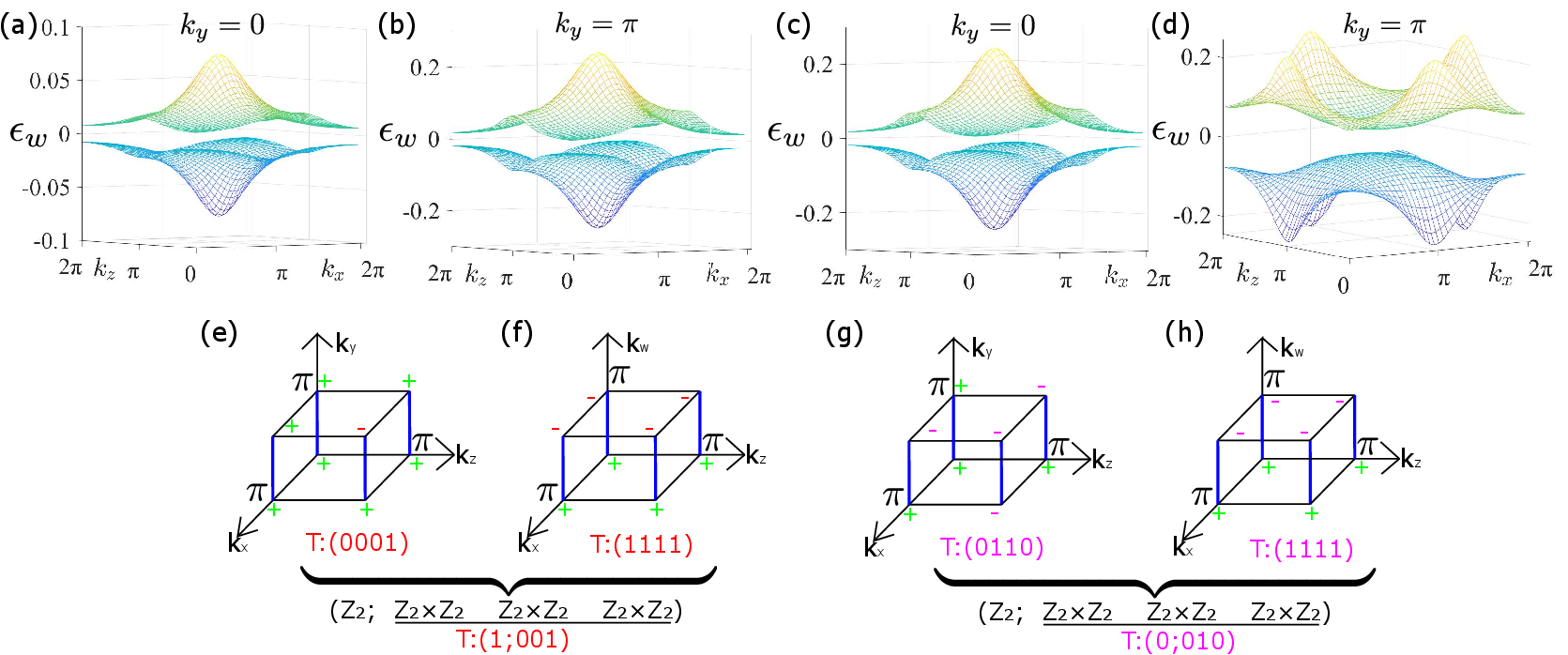}
    \caption{[(a) and (b)] Typical Wannier band structure in phase $T_1^{\text{II}}$. 
    [(c) and (d)] Typical Wannier band structure in phase $T_2^{\text{II}}$. (e) and (f) The topological invariants in phase $T_1^{\text{II}}$. 
   [(g) and (h)] The topological invariants in phase $T_2^{\text{II}}$. Parameters are: $m_1=2.5,m_2=0.5$ in phase $T_1^{\text{II}}$, $m_1=m_2=0.5$ in phase $T_2^{\text{II}}$.
   }
    \label{figsuppcharacterize2}
\end{figure}

\subsection{Model III: 6-dimensional second-order topological insulator}
\label{6d2omodelchap}
We consider the following 6D lattice model:
\begin{equation}
    \begin{split}
H(\mathbf{k})=&\sin k_x\Gamma_1+\sin k_y\Gamma_2+\sin k_z\Gamma_3+\sin k_w\Gamma_4+\sin k_l\Gamma_5+\sin k_f\Gamma_6\\
&+(m_1+\cos k_x+\cos k_l)\Gamma_7+(m_2+\cos k_y+\cos k_z+\cos k_w+\cos k_f)\Gamma_8
    \end{split}
    \label{6d2omodel}
\end{equation}
Where 
$\Gamma_{(1)-(9)}=(\tau_z\sigma_zs_z\beta_z,\tau_y\sigma_zs_z\beta_z,\tau_x\sigma_zs_z\beta_z,\sigma_ys_z\beta_z,s_y\beta_z,\beta_y,s_x\beta_z,\beta_x,\sigma_xs_z\beta_z)$. The Hamiltonian has rotational symmetry: $C_4^{x,l}=\frac{I-\Gamma_4\Gamma_5}{2}$, $C_4^{x,l}H(k_x,k_l,k_y,k_z,k_w,k_f)(C_4^{x,l})^{-1}=H(-k_l,k_x,k_y,k_z,k_w,k_f)$, and $C_4^{y,z}$, $C_4^{y,w}$, $C_4^{y,f}$, $C_4^{z,w}$, $C_4^{z,f}$, $C_4^{w,f}$ are defined in a similar way. It has reflection symmetry: $M_x=\Gamma_1\Gamma_9$, $M_xH(k_x,k_y,k_z,k_w,k_l,k_f)M_x=H(-k_x,k_y,k_z,k_w,k_l,k_f)$, and $M_y,\;M_z,\;M_w,\;M_l,\;M_f$ symmetry can be defined in a similar way. It also has time-reversal symmetry: $T=\tau_yK$ or $T=\sigma_yK$ with $TH(\mathbf{k})T^{-1}=H({-\mathbf{k}})$

We consider the case with open (periodic) boundary conditions along $l,f$ {($x,y,z,w$)} directions. We solve for the corner states at $l=0,f=0$, following the method discussed above, we get the projection operators for the spin states: $P_{l\leftarrow}=\frac{I+i\Gamma_5\Gamma_7}{2}=\frac{I+s_z}{2},\;P_{f\downarrow}=\frac{I+i\Gamma_6\Gamma_8}{2}=\frac{I+\beta_z}{2}$. 

$\bullet$ We first discuss the case when $-2<m_1<0,\;-4<m_2<-2$, and find one effective Dirac cone lying in the corner, around the momentum $(G_x,G_y,G_z,G_w)=(0,0,0,0)$, so the effective Hamiltonian of corner state is:
\begin{equation}
H_\text{corner}=k_x\sigma_z\tau_z+k_y\sigma_z\tau_y+k_z\sigma_z\tau_x+k_w\sigma_y
\end{equation}
The corner energy gap can be opened by adding proper mass terms
in the forms of $\Gamma_9$, $i\Gamma_5\Gamma_7\Gamma_9$, $\Gamma_6\Gamma_8\Gamma_9$, $\Gamma_1\Gamma_2\Gamma_3\Gamma_4$ or their combinations, then HIQHE emerges at the corner. 
We consider the mass term $H_\text{M}=-M\Gamma_1\Gamma_2\Gamma_3\Gamma_4=M\sigma_x$ for example, then in the presence of weak magnetic field $B_{zy}$ perpendicular to $zy$ plane 
the low energy effective corner Hamiltonian becomes
\begin{equation}
    H=\sigma_z\{k_x\tau_z+[k_y+\frac{B_{zy}z}{2}]\tau_y+[k_z-\frac{B_{zy}y}{2}]\tau_x\}+k_w\sigma_y+M\sigma_x
\end{equation}
When a weak electric field $E_x$ is applied,
using linear response theory (Kubo formula) with expansion up to second-order case~\cite{dimensionalreduction} or using the semi-classical equations of motion for a Bloch electron in weak electric and magnetic fields~\cite{coldatom4dhall}, we can get the 4D Hall-type electric current response {of the corner state}:
\begin{equation}
    \vec{j}=(j_x,j_y,j_z,j_w)=(0,0,0,-\frac{C_2}{4\pi^2}B_{zy}E_x).
\end{equation}
where 
\begin{equation}
    C_2=\frac{1}{32\pi^2}\int d^4k\epsilon_{i_1,i_2,i_3,i_4}\text{Tr}[\Omega_{i_1i_2}\Omega_{i_3i_4}]=\frac{1}{2}.
\end{equation}
with $(i_1,i_2,i_3,i_4)\in(x,y,z,w)$

Alternatively, we can derive such 4D Hall-type electric current response
directly in a simpler way. 
We define $\Pi$ as the dynamical operator satisfying commutation relation $[\Pi,\Pi^{\dagger}]=2B_{zy}$, that is:
\begin{equation}
    \Pi=k_z-ik_y+\frac{1}{2}B_{zy}(-y-iz),\Pi^{\dagger}=k_z+ik_y+\frac{1}{2}B_{zy}(-y+iz),
\end{equation}
then the effective Hamiltonian for $E_x=0$ can be written as:
\begin{equation}
    \begin{split}
		H={\left[\begin{array}{cccc}
				k_x&\Pi\\
                \Pi^{\dagger}&-k_x\\
			\end{array}
			\right]}\otimes\sigma_z+k_wI_2\otimes\sigma_y+MI_2\otimes\sigma_x.
	\end{split}
\end{equation}
After second quantization: $\Pi=\sqrt{2|B_{zy}|}a$ (or $\Pi^{\dagger}=\sqrt{2|B_{zy}|}a$) for $B_{zy}>0$ (or $B_{zy}<0$), we can reduce the effective Hamiltonian to 2D case:
\begin{equation}
	\begin{cases}
		H_{\pm n}=\pm\sqrt{k_x^2+\cdot 2B_{zy}n}\sigma_z+k_w\sigma_y+M\sigma_x\;(n>0)\\
		H_{0}=-\text{sign}[B_{zy}]k_x\sigma_z+k_w\sigma_y+M\sigma_x\\
	\end{cases}
\end{equation}
The Berry phase of $H_{+n}$ is opposite to $H_{-n}$. When $E_x$ is applied, the Hall-type response of $H_{+n}$ and $H_{-n}$ can be cancelled out, and $H_{0}$ leads to half-quantized Hall current along $w$: $j_w=-\frac{1}{2}\frac{1}{4\pi^2}B_{zy}E_x$, notice that we have
multiplied the filling factor $\frac{1}{2\pi}B_{zy}$ to obtain the current density.

$\bullet$ For parameter region $-2<m_1,m_2<0$, we solve for the corner states and find three Dirac cones at each corner, around the momenta $(0,0,0,\pi)$, $(0,0,\pi,0)$, $(0,\pi,0,0)$. When $H_\text{M}=M\sigma_x$ is applied on the corner sites to open the gaps, we can get the $\frac{3}{2}$-quantized Hall-type response for each corner:
\begin{equation}
    \vec{j}=(j_x,j_y,j_z,j_w)=(0,0,0,\frac{3}{2}\frac{1}{4\pi^2}B_{zy}E_x)
\end{equation}
Generally, if we add electromagnetic field in the 4D corner state, after doing tedious calculation by the method above, we can get the 4D Hall-type current response:
\begin{equation}
    j_{i_1}=\frac{N_\text{Dirac}^+-N_\text{Dirac}^-}{2}\frac{1}{2\cdot(2\pi)^2}\epsilon_{i_1i_2i_3i_4}B_{i_2i_3}E_{i_4}
\end{equation}
Where $N_\text{Dirac}^+$ ($N_\text{Dirac}^-$) is the total number of massive Dirac cones with positive (negative) effective mass such that each contributes to $+\frac{1}{2}$ ($-\frac{1}{2}$) 4D Hall current.

We have discussed the properties of a single corner, the physics for other corners are similar. Adding all the contributions of corners, we can get the total Hall current determined by the total second Chern number. Additionally, we can calculate layer-resolved second Chern numbers which further reflect local Chern-Simons contributions of Hall-type response of the corner states according to  Eq.~\ref{locallocalhigh} in Sec.~\ref{layer2nd}.  
Also, the corner states may appear for open boundaries along directions: $(l,y)$, or $(l,z)$, or $(l,w)$, or $(x,y)$, or $(x,z)$, or $(x,w)$, or $(x,f)$ with periodic boundary conditions along the rest directions.

\section{Experimental consideration}
We first describe the generalized Streda formula in Sec.~\ref{pdpb}, which connect the layer-resolved Chern number with 
variation of mode density 
as a function of magnetic field.
Then we describe the probing method to detect mode density via photonic input-output relation in Sec.~\ref{detect4d4d}. 

\subsection{Layer-resolved Chern number and Streda formula}
\label{pdpb}
The Streda formula~\cite{streda}, characterizes the topological magneto-electric response, is well known. It relates the Chern number to the change in the density of a system as an applied magnetic field is varied, that is: $C=2\pi\frac{\partial\rho_\text{total}}{\partial B_{xz}}$, where $\rho_\text{total}$ is total density of occupied bands, $B_{xz}$ is the weak magnetic magnetic field.
Here we generalized the Streda formula to detect the layer-resolved Chern number in our system. First, we consider the 4D lattice model Eq.~(1) in the main text, then we discuss the general cases.

For our model Eq.~(1) in the main text, we consider open (periodic) boundary conditions along $yw$ (xz) directions, and will prove that 
the layer-resolved Chern number $C(y,w)$ at site $(y,w)$ satisfies $C(y,w)=2\pi\frac{\partial \rho(y,w)}{\partial B_{xz}}$, where $\rho(y,w)$ is the mode density at the lattice site $(y,w)$, $B_{xz}$ is the applied weak magnetic field perpendicular to $xz$ plane. The current conservation equation is as follows
\begin{equation}
    \sum_{q=x,z}\partial_{q}j^q(y,w)+\partial_t\rho(y,w)=0
    \label{rbiden1}
\end{equation}
Where $j^q(y,w)$ is the local current, and $\rho(y,w)$ is mode density. According to the Kubo formula and layer-resolved local Chern number discussed in Sec.~\ref{layerresolvedchern}, we have
\begin{equation}
    j^q(y,w)=\frac{C(y,w)}{2\pi}\epsilon_{qq'}E_{q'}
    \label{rbiden2}
\end{equation}
The Maxwell equation
\begin{equation}
    \partial_tB_{qq'}+\epsilon_{qq'}\partial_qE_{q'}=0
    \label{rbiden3}
\end{equation}
According to Eq.~\ref{rbiden1}, Eq.~\ref{rbiden2}, Eq.~\ref{rbiden3}, we can get the generalized Streda formula:
\begin{equation}
    C(y,w)=2\pi\frac{\partial \rho(y,w)}{\partial B_{xz}}
\end{equation}
The Streda formula can be generalized to higher-dimensional and higher-order cases. Following the discussions above and in Sec.~\ref{layer2nd}, we can get the variation of mode density $\rho(q_1,q_2,\cdot\cdot\cdot,q_n)$ as functions of magnetic field
\begin{equation}
    \rho(q_1,q_2,\cdot\cdot\cdot,q_n,\mathbf{B})-\rho(q_1,q_2,\cdot\cdot\cdot,q_n,\mathbf{0})=\frac{2\pi}{(2\pi)^{(d-n)/2}}\frac{C(q_1,q_2,\cdot\cdot\cdot ,q_n)}{2^{(d-n)/2}}\epsilon_{i_1i_2\cdot\cdot\cdot i_{d-n}}B_{i_1i_2}B_{i_3i_4}\cdot\cdot\cdot B_{i_{d-n-1}i_{d-n}}
    \label{rholocalhigh}
\end{equation}
Where $(i_1,i_2,\cdot\cdot\cdot,i_{d-n})\in(q_{n+1},q_{n+2},\cdot\cdot\cdot,q_d)$.

\begin{figure}
    \centering
    \includegraphics[width=0.8\linewidth]{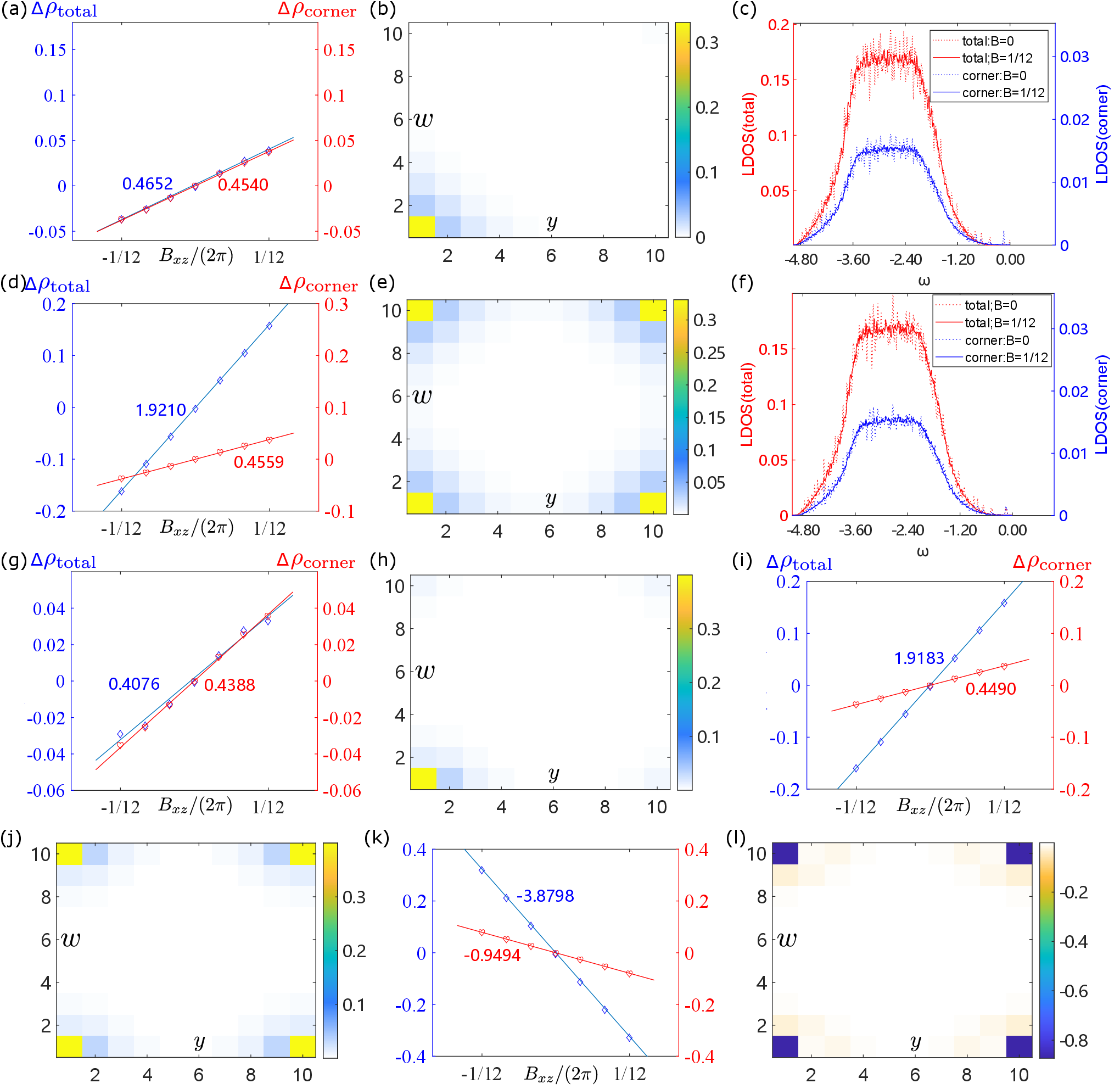}
    \caption{Numerical simulation of the mode density through the transmission. (a)-(c) Results for the model Eq.~(1) in the main text with $m_1=m_2=1.5$. The mass term $H_\text{M}=M\cdot i\Gamma_1\Gamma_4=-Ms_z$ is applied on one corner. The slopes (marked by the digits) in (a) represent the corresponding Chern number. $\Delta\rho_\text{total}=\sum_{y_o,w_o} [\rho(y_0,w_0,B_{xz})-\rho(y_0,w_0,0)]$ is the total mode-density derivation (diamonds), similarly for the corner mode-density derivation (hearts) $\Delta\rho_\text{corner}$, except the summation is taken over a single corner region. The extracted layer resolved Chern number is shown in (b). Local density of states with and without magnetic field are shown in (c).
    (d)-(e) Same as that in (a)-(c) except that the mass term $H_\text{M}=M\cdot i\Gamma_1\Gamma_4=-Ms_z$ is applied on four corners.
    [(g) and (h)] Same as that in (a,b) but for the model II Eq.~(\ref{1add3model}) with $m_1=2.5,m_2=0.5$. 
    [(i) and (j)] Same as that in (d,e) but for the model II Eq.~(\ref{1add3model}) with $m_1=2.5,m_2=0.5$. 
    [(k) and (l)] Same as that in (d,e) but for the model II Eq.~(\ref{1add3model}). In (a)-(l) are: $\gamma=0.01,M=0.4$.}
    \label{detectrho}
\end{figure}

\subsection{Detecting Method}
\label{detect4d4d}
Here we show how the photonic transmission spectra~\cite{luozhou2015,localhubbard} can be used to detect the mode density and thereby  the local Chern number.
We consider each cavity has two input-output ports, we pump the cavity from one port, and measure the transmission from the other port.
The photonic transport property is described by the input-output formalism in the Heisenberg picture~\cite{luozhou2015,localhubbard}, that is:
\begin{equation}
    \frac{dc_{\mathbf{r}}(t)}{dt}=\frac{1}{i\hbar}\sum_{\mathbf{r}'}H_{\mathbf{r}\mathbf{r}'}c_{\mathbf{r}'}(t)-\gamma c_{\mathbf{r}}(t)-\sqrt{\gamma}a_{\text{in},\mathbf{r}}(t)
    \label{inout11}
\end{equation}
\begin{equation}
    a_{\text{out},\mathbf{r}}(t)=a_{\text{in},\mathbf{r}}(t)+\sqrt{\gamma}c_\mathbf{r}(t)
    \label{inout12}
\end{equation}
\begin{equation}
    b_{\text{out},\mathbf{r}}(t)=\sqrt{\gamma}c_\mathbf{r}(t)
    \label{inout13}
\end{equation}
Where $c_\mathbf{r}$ is the cavity mode with  $\mathbf{r}=(x,y,z,w,\alpha)$ the site and spin indices, $a$ and $b$ represents the input-output fields at the two ports. {$\gamma$ is input-output coupling coefficient,} 
$H_{\mathbf{r}\mathbf{r}'}$ is the real-space Hamiltonian matrix.
According to Eq.~\ref{inout11}, Eq.~\ref{inout12}, Eq.~\ref{inout13}, we can get:
\begin{equation}
    c_{\text{out},\mathbf{r}}=\sum_{\mathbf{r}'}\left[\frac{-i\gamma}{\omega-H+i\gamma}\right]_{\mathbf{r}\mathbf{r}'}b_{\text{in},\mathbf{r}'}.
    \label{coutbin}
\end{equation}
We can define $\mathbb{T}_{\mathbf{r}_\text{o};\mathbf{r}_\text{i}}(\omega)=\langle \mathbf{r}_\text{o}|\frac{-i\gamma}{\omega-H+i\gamma}|\mathbf{r}_\text{i}\rangle$ so that
\begin{equation}
c_{\text{out},\mathbf{r}_\text{o}}=\mathbb{T}_{\mathbf{r}_\text{o};\mathbf{r}_\text{i}}(\omega)b_{\text{in},\mathbf{r}_\text{i}}
\end{equation}
Noticing $H=\sum_m E_m|u_m\rangle\langle u_m|$, we can obtain
\begin{equation}
    \mathbb{T}_{\mathbf{r}_\text{o};\mathbf{r}_\text{i}}(\omega)=\int\frac{dk_xdk_z}{(2\pi)^2}\sum_m\frac{-i\gamma u_m(y_\text{o},w_\text{o},\alpha_\text{o},k_x,k_z)u_m^*(y_\text{i},w_\text{i},\alpha_i,k_x,k_z)}{\omega-E_{m}(k_xk_z)+i\gamma}e^{ik_x(x_\text{o}-x_\text{i})}e^{ik_z(z_\text{o}-z_\text{i})}
\end{equation}

In $xz$ space, we fix the input pumping site position (e.g., at the center) and
pump all the sites in the $yw$ space in turns, along with all the pseudo-spin component. Then, for a given output site in the $yw$ space, we measure the total intensity transmission, and we can obtain
\begin{equation}
    \mathcal{T}_{y_\text{o}w_\text{o}}(\omega)=\sum_{\mathbf{r}'_\text{o},\mathbf{r}''_\text{i}}|\mathbb{T}_{\mathbf{r}_\text{o},\mathbf{r}_\text{i}}(\omega)|^2
\end{equation}
with 
$\mathbf{r}'_\text{o}=(x_\text{o},z_\text{o},\alpha_\text{o}),\mathbf{r}''_\text{i}=(y_\text{i},w_\text{i},\alpha_\text{i})$.
We consider that $\gamma$ is smaller compared to the bandwidth and gap,
so that we can use the approximation $\gamma\rightarrow0^+$ and obtain the local density of states
\begin{equation}
    \rho_{y_\text{o},w_\text{o}}(\omega)=\lim_{\gamma\rightarrow0^+}\frac{1}{\pi\gamma}\sum_{\mathbf{r}'_\text{o}\mathbf{r}'_\text{i}}|\mathbb{T}_{\mathbf{r}_\text{o},\mathbf{r}_\text{i}}(\omega)|^2
\end{equation}
We have used that
\begin{equation}
    \rho_{y_\text{o},w_\text{o}}(\omega)=\sum_{k_x,k_z,\alpha}\sum_{E_m(k_x,k_z)<0}u_{m}^*(y_\text{o},w_\text{o},\alpha,k_x,k_z)u_m(y_\text{o},w_\text{o},\alpha,k_x,k_z)\delta(E_m-\omega)
    \label{rhoy0w0}
\end{equation}
By summing over all the output sites $y_\text{o},w_\text{o}$, we can get the total density of states.
\begin{equation}
    \rho_{\text{total}}(\omega)=\sum_{y_\text{o},w_\text{o}}\rho_{y_\text{o},w_\text{o}}(\omega)
\end{equation}
We scan the frequency of the input laser to cover the valence band, then we can get the target mode density $\rho(y_\text{o},w_\text{o})$:
\begin{equation}
    \int_{-\infty}^0d\omega\rho_{y_\text{o},w_\text{o}}(\omega)=\lim_{\gamma\rightarrow0^+}\frac{1}{\pi\gamma}\int_{-\infty}^0d\omega \mathcal{T}_{y_\text{o}w_\text{o}}(\omega)=\rho(y_\text{o},w_\text{o})
\end{equation}
Notice that
\begin{equation}
    \rho(y_\text{o},w_\text{o})=\sum_{k_x,k_z,\alpha}\sum_{E_{m}(k_x,k_z)<0}u_{m}^*(y_\text{o},w_\text{o},\alpha,k_x,k_z)u_{m}(y_\text{o},w_\text{o},\alpha,k_x,k_z)
    \label{rhoy0w0_2}
\end{equation}

\end{widetext}

\end{document}